\documentclass[aps,prd,twocolumn,nofootinbib,superscriptaddress,amsmath,amssymb]
{revtex4-2}
\usepackage{CJK}                     
\usepackage{graphicx}                
\usepackage{mathptmx}                
\usepackage{bm}
\usepackage{physics}
\usepackage{dcolumn}                 
\usepackage{multirow}                
\usepackage{xcolor}
\usepackage{hyperref}
\hypersetup{
breaklinks=true,colorlinks=true,
linkcolor=blue,citecolor=blue,filecolor=magenta,urlcolor=black}

\long\def\OFF#1{}

\def\be{\begin{equation}} \def\ee{\end{equation}}
\def\bal#1\eal{\begin{align}#1\end{align}}
\def\non{\nonumber}

\def\eps{\varepsilon}
\def\phi{\varphi}
\def\la{\lambda}

\def\om{\omega}
\def\ms{\,M_\odot}

\def\km{\,\text{km}}
\def\fm3{\,\text{fm}^{-3}}

\def\xp{x_p}
\def\khz{\,\text{kHz}}

\def\nm{\text{N}}
\def\dm{\text{D}}
\def\fdm{F_{\rm D}}
\def\rstar{r_*}

\def\fgl{f_{\rm gl}}
\def\frel{f_{\rm rel}}

\begin{document}

\title{Two-fluid \texorpdfstring{$f$}{f}-mode oscillations
of dark-matter-admixed neutron stars}

\begin{CJK*}{UTF8}{gbsn}

\newcommand{\ccnu}
{Institute of Astrophysics, Central China Normal University,
Luoyu Road 152, Wuhan 430079, China}
\newcommand{\hue}
{School of Optoelectronic Information Engineering,
Institute of Astronomy and High Energy Physics,\\
Hubei University of Education,
Second Gaoxin Road 129, Wuhan 430205, China}
\newcommand{\cug}
{School of Mathematics and Physics, China University of Geosciences,
Lumo Road 388, Wuhan 430074, China}
\newcommand{\hust}
{School of Physics, Huazhong University of Science and Technology,
Luoyu Road 1037, Wuhan 430074, China}
\newcommand{\infn}
{INFN Sezione di Catania, Dipartimento di Fisica,
Universit\'a di Catania, Via Santa Sofia 64, 95123 Catania, Italy}

\author{Zi-Yue Zheng (郑子岳)}\affiliation{\ccnu}
\author{Ting-Ting Sun (孙婷婷)}\affiliation{\hue}
\author{\hbox{Huan Chen (陈欢)}}\affiliation{\cug}
\author{\hbox{Xiao-Ping Zheng (郑小平)}}
\email{Email:zhxp@ccnu.edu.cn}\affiliation{\ccnu}\affiliation{\hust}
\author{\hbox{Jin-Biao Wei (魏金标)}}\affiliation{\cug}
\author{G. F. Burgio}\affiliation{\infn}
\author{H.-J. Schulze}\affiliation{\infn}

\begin{abstract}
We study quadrupolar $f$-mode oscillations
of dark-matter-admixed neutron stars (DANSs)
in full general relativity (GR).
The ordinary component is described by microscopic Brueckner-Hartree-Fock matter
matched to the Shen2020 crust,
while the dark matter (DM) component is treated
as a cold self-interacting fermion fluid coupled to ordinary matter only by gravity.
For fixed-DM-fraction sequences we solve the polar two-fluid perturbation equations
with an outgoing gravitational-wave (GW) boundary condition,
obtaining complex eigenfrequencies rather than only real mode frequencies.
The spectrum contains two principal $f$-like sequences.
Their local character can be ordinary-matter-led, DM-led, or mixed,
and is diagnosed using the component kinetic energies,
the displacement overlap,
and the cancellation of the matter quadrupole.
A main result is that, for intermediate DM fractions,
one of the two-fluid branches can become weakly radiating,
with damping times enhanced by several orders of magnitude.
The same calculation gives the outgoing Zerilli amplitude
and the GW damping time,
which we use to estimate the GW energy required
to reach a prescribed detector threshold.
Thus the analysis extends previous two-fluid Cowling studies
by retaining metric perturbations and the radiative boundary condition.
\end{abstract}

\maketitle
\end{CJK*}

\section{Introduction}

Neutron stars (NSs) give access to cold dense matter
at densities of several times nuclear saturation density.
Their masses, radii, and tidal deformabilities are
tied to the equation of state (EOS) in a regime
that cannot be reproduced in terrestrial experiments.
Measurements of massive pulsars
\cite{Antoniadis13,Fonseca21,Romani22},
X-ray pulse-profile modeling
\cite{Riley19,Miller19,Riley21,Miller21,Vinciguerra24,Dittmann24},
and gravitational-wave (GW) observations of binary NS mergers
\cite{Abbott17a,Abbott18,Pang21,Raaijmakers21,Rutherford24}
have already ruled out broad classes of soft EOSs.
These observables,
however,
do not by themselves identify the matter content of the star.
EOS softening, hadron-quark conversion,
or an additional matter component
can lead to similar mass-radius curves
and comparable tidal responses.
This degeneracy is one reason to look beyond static observables.
Oscillation modes respond not only to the equilibrium profile,
but also to the degrees of freedom
that are able to move inside the star.

Non-radial oscillations (NROs) provide such a probe.
For multipoles with $l\geq2$,
stellar pulsations can radiate GWs
and carry information from the stellar interior \cite{Thorne67}.
They can be excited in core collapse,
in binary-merger remnants,
by dynamical tides in eccentric or spinning binaries,
or by disturbances in isolated stars
\cite{Radice19,Stergioulas11,Vretinaris20,Soultanis22,Doneva13,Chirenti17,Steinhoff21,Pratten20}.
For a nonrotating star,
the perturbations separate into axial and polar sectors.
The polar sector contains the pulsational fluid motion
and couples directly to GW emission.
It includes the pressure $p$ modes,
the buoyancy-driven $g$ modes,
the fundamental $f$ mode,
and spacetime-dominated $w$ modes
\cite{Thorne67,Price69,Lindblom83,Detweiler85,Andersson96,Kokkotas99,Kokkotas01}.
The $g$ modes are sensitive to stable stratification,
composition gradients,
density discontinuities,
and phase structure
\cite{Zhao22a,Constantinou21,Zhao22b,Zheng23}.
The $p$ modes occur at higher frequencies
and are mainly governed by compressibility \cite{Kokkotas99,Kokkotas01}.
The $f$ mode is a global oscillation
with a typical frequency $f_f=\Re\omega_f/2\pi\sim1.3$--$3\khz$,
lying between the low-frequency $g$ modes
and the higher-frequency $p$ modes \cite{Kokkotas99,Kunjipurayil22}.
Its frequency and damping time
are strongly correlated with the mean density,
compactness, and tidal deformability of ordinary NSs
\cite{Andersson98,Kokkotas99,Chan14,Sotani21,Zhao22b,Pradhan22}.
This makes the $f$ mode one of the most useful targets
for GW asteroseismology.

For ordinary single-fluid NSs,
the full NRO problem in general relativity (GR) is well established.
In the Lindblom-Detweiler formulation,
the fluid displacement and metric perturbations
are evolved together inside the star,
matched to an exterior Zerilli solution,
and constrained by a purely outgoing wave condition at infinity
\cite{Thorne67,Lindblom83,Detweiler85,Zhao22b}.
The eigenfrequency is then complex.
Its real part gives the oscillation frequency,
and its imaginary part gives the GW damping time.
Full-GR calculations of the $f$ and $p$ modes
have quantified how the corresponding frequencies
and damping times depend on compactness, tidal deformability,
and the EOS \cite{Kunjipurayil22,Sotani21,Zhao22b,Pradhan22,Zheng25a}.
The low-frequency $g$ modes probe a different part of the physics.
They are tied to buoyancy from composition gradients,
density discontinuities, or phase structure,
and have been studied in hadronic and hybrid-star models
\cite{Zhao22a,Constantinou21,Zhao22b,Zheng23}.
In the present work,
we focus on the quadrupolar $f$ mode.
For this mode,
an outgoing-wave calculation is required
once the damping time or the strain amplitude is part of the analysis.
The damping time sets the lifetime of the ringdown,
whereas the strain amplitude depends on
the conversion of oscillation energy into outgoing GWs.

Dark matter (DM) provides a physically motivated source
of an additional stellar component.
The evidence for DM is gravitational
on galactic and cosmological scales,
but its particle mass and interactions remain unknown.
Compact stars therefore provide
a complementary high-density environment
in which dark-sector physics may leave structural or dynamical signatures
\cite{Baryakhtar22,Bramante24}.
Because NSs are compact and baryon rich,
they may capture or accumulate DM during their evolution
\cite{Press85,Krauss86,Gould87,Gould88,Bramante14,Bell21,Baryakhtar22,Bramante24}.
The amount of accumulated DM is controlled
by the ambient DM density
and the relevant scattering cross sections
\cite{Press85,Gould87,Gould88,Bramante14,Bell21,Bramante24}.
It also depends on the stellar formation history,
including whether the DM component is captured after birth,
inherited from the progenitor environment,
or produced during core collapse \cite{Baryakhtar22,Bramante24}.
These ingredients are highly model dependent
\cite{Baryakhtar22,Bramante24}.
For standard capture in an ordinary Galactic environment,
the resulting DM fraction is expected to be small
\cite{Ellis18,Bramante24}.
Larger fractions therefore require special environments,
additional dark-sector dynamics,
or a primordial admixture
\cite{Bramante14,Tulin13,Kouvaris15,Leung11,Li12,Kain21,Bramante24}.
They may also arise from conversion mechanisms,
such as neutron decays into dark-sector particles
\cite{Fornal18,McKeen18,Berryman22,Shirke23}.
The formation or capture history of the DM component is not modeled here.
Instead,
the DM mass fraction is treated as a controlled macroscopic parameter.
This choice follows the common practice
in Dark-matter-admixed neutron star (DANS) studies
and allows the two-fluid dynamical effects
to be isolated from the uncertain formation problem.

DANSs provide a concrete example
of the degeneracy of static observables.
If the dark particles are stable on stellar time scales
and interact with ordinary matter mainly through gravity,
the equilibrium configuration is described
by two gravitationally coupled fluids \cite{Leung11,Li12,Kain21}.
The ordinary-matter and DM components have separate EOSs,
pressures, conserved currents, and radial extents,
while the spacetime is sourced by their total stress energy.
Depending on the DM particle mass,
self-interaction strength,
and total abundance,
the DM component may form a compact core
or an extended halo outside the visible ordinary-matter surface
\cite{Sandin09,Narain06,Tulin13,Kouvaris15,Maselli17,Nelson19,Hippert22,Shawqi24,Liu23,Liu24,Wei26}.
DM can modify mass-radius relations,
tidal deformabilities,
moments of inertia,
cooling behavior, and stability properties
\cite{Leung11,Li12,Tolos15,Ellis18,Quddus20,Das20,Yang21,Leung22,Dengler22,Gleason22,Caballero24,Kumar25a,Kumar25b,Biesdorf25,Routaray25,Zhou25,Wei26}.
Radial oscillations and stability analyses
have further shown that the stable parameter space
of a two-fluid star cannot always be inferred
from either component EOS separately
\cite{Kain21,Gleason22,Caballero24,Kumar25b}.
The same total mass may then correspond to different visible radii,
core-halo structures,
and relative fluid contributions.
In this situation,
oscillation spectra can provide information
that is not accessible from static observables alone.

The perturbation problem of a DANS cannot be obtained
by merely replacing the EOS in the single-fluid equations.
Ordinary matter and DM carry independent displacement fields,
and the two perturbed fluids are coupled through the metric
and the common gravitational potential.
The situation is conceptually related to multifluid descriptions of superfluid NSs,
where relative motion between superfluid neutrons
and the charged component can introduce additional mode families
\cite{Baym69,Epstein88,Lombardo01,Lindblom94,Comer99,Andersson02,Chamel08,Lin08}.
DANSs differ in an important way.
Here the two components are not assumed
to have entrainment or direct nongravitational coupling,
and they may terminate at different radii
\cite{Sandin09,Leung11,Kain21,Caballero24,Kumar26}.
In the mirror-DM core configurations of Ref.~\cite{Kumar26},
the two nodeless $f$-like continuations
were classified from their radial eigenfunctions
as outer- and inner-fluid-led branches.
These correspond naturally to the global-like
and relative continuations followed in the present work.
The co-moving or counter-moving character
instead describes the local relative phase
of the two displacement fields;
this terminology originates in
relativistic two-fluid models
of superfluid NSs
\cite{Lindblom94,Comer99,Andersson02}.
For GW emission,
it is also relevant how the fluid motion is
distributed between the two components.
In the two-fluid Einstein equations,
the metric perturbations are sourced
by the sum of the component stress-energy perturbations \cite{Kumar26}.
At leading mass-quadrupole order,
the corresponding component quadrupoles add linearly
and can therefore interfere constructively
or partially cancel \cite{Thorne80}.
Whether such source-level interference
produces a weakly radiating normal mode
must be established from the full outgoing-wave problem.

The relativistic Cowling approximation \cite{Cowling41},
which neglects metric perturbations,
has enabled first surveys of DANS oscillations.
Calculations based on effective single-fluid treatments
or Cowling perturbations showed that the DM
component can shift the fluid-mode spectrum
\cite{Flores24,Shirke24,Thakur24,Shirke25}.
In particular,
a recent Cowling study of DANS $g$ modes
found that the frequency shift
is mainly controlled by the DM fraction,
and emphasized the degeneracy
between DM admixture and other composition effects \cite{Shirke25}.
A two-fluid Cowling formulation
for self-interacting fermionic DM,
with ordinary matter and DM coupled only through gravity,
then identified component-associated oscillation frequencies \cite{Sotani25a}.
That study used a fixed central-density ratio and a single ordinary-matter EOS.
The same framework was later applied to fixed-DM-fraction sequences
with several ordinary-matter EOSs,
mainly to examine whether familiar $f$-mode universal relations
survive in DANSs \cite{Sotani25b}.
In these gravity-only models,
however,
neglecting metric perturbations removes the direct coupling mediated
by the perturbed spacetime.
The Cowling equations are therefore solved
as component-associated fluid problems
on the same two-fluid background.
They provide useful approximations to the frequency trends
of ordinary- or DM-led fluid motion,
but they do not determine the coupled radiative normal modes,
the radiative character of the counter-moving branch,
the GW damping time,
or theoutgoing radiative amplitude.

The full polar perturbation equations
for gravitationally coupled two-fluid stars
have recently been formulated in full GR \cite{Kumar26}.
That work starts from independently conserved perfect-fluid currents
coupled only through the common spacetime,
derives the even-parity Einstein--Euler perturbation equations,
and extends the Lindblom-Detweiler variables
by introducing separate displacement and pressure variables for each fluid.
The metric perturbations are sourced by the sum of the ordinary-matter
and DM stress-energy perturbations.
The formulation also specifies the regularity conditions at the center,
the Lagrangian-pressure boundary condition at each component surface,
and the matching to an exterior Zerilli--Regge-Wheeler solution
with an outgoing-wave condition.
As a first application,
Ref.~\cite{Kumar26} computed representative real-frequency spectra
for mirror-DM two-fluid models
and used the eigenfunctions and node structure
to identify inner- and outer-fluid-led $f$ and $p$ branches.
The imaginary part of the eigenfrequency,
the GW damping time, and the outgoing radiative amplitude
were not extracted there.
That first application was restricted
to real-frequency residual spectra
and did not extract the imaginary part of the eigenfrequency,
the GW damping time, or the asymptotic radiative amplitude.
In the present work,
We also compute the GW damping time $\tau$
and use the mode frequencies and damping times
to estimate the GW energy required
to reach a prescribed detector threshold.
This allows us to assess
the observational relevance of the different radiative segments.

Here we use this full-GR two-fluid framework
to compute the radiative $f$-like oscillations of DANSs.
The ordinary component is described
by microscopic Brueckner-Hartree-Fock (BHF) matter
matched to the Shen2020 crust.
The DM component is modeled as a cold,
self-interacting fermion fluid.
We construct sequences at fixed DM mass fraction
and solve the polar eigenvalue problem
with outgoing GW boundary conditions.
Before universal relations for such systems can be assessed,
the relevant radiative two-fluid branches must be identified.
Here we first isolate the two principal two-fluid $f$-like sequences,
follow their evolution with DM
fraction and core-halo structure,
and distinguish ordinary-matter-led, DM-led,
and mixed oscillations locally.
We introduce diagnostics based on the component kinetic-energy content,
the displacement overlap,
and the cancellation of the matter quadrupole.
We also compute the GW damping time $\tau$
and estimate the strain amplitude for a prescribed total GW energy,
which allows us to discuss the observational relevance
of the different radiative segments.

This article is organized as follows.
In Sec.~\ref{s:eos},
we describe the ordinary-matter and DM EOSs.
In Sec.~\ref{s:osc},
we formulate the two-fluid equilibrium problem,
the polar perturbation equations,
and the numerical eigenvalue method.
The numerical results are presented in Sec.~\ref{s:res}.
Their physical implications are discussed in Sec.~\ref{s:disc}.
We summarize our conclusions in Sec.~\ref{s:end}.
Geometrized units $G=c=1$ are adopted throughout the article.

\section{Equation of state}
\label{s:eos}

\subsection{Ordinary matter}

We construct the EOS of ordinary matter
from microscopic BHF calculations for uniform nuclear matter (NM),
joined at low density to a crust EOS.
In this framework,
short-range correlations from the bare nucleon-nucleon interaction
are resummed into the in-medium reaction matrix $K$.
For a baryon density $\rho$ and proton fraction $\xp\equiv\rho_p/\rho$,
$K$ satisfies the Bethe-Goldstone equation
\be
 K(\rho,\xp;E) = V + \Re\sum_{1,2}V
 \frac{\ket{12} Q \bra{12}}{E-e_1-e_2} K(\rho,\xp;E) \:
\label{e:k}
\ee
and determines the single-particle potential through
\be
 U_1(\rho,\xp) = \Re\sum_{2<k_F^{(2)}}
 \expval{K(\rho,\xp;e_1+e_2)}{12}_a \:,
\label{eq:uk}
\ee
where $V$ is the two-nucleon interaction,
$E$ is the starting energy,
and $Q$ projects intermediate states above the Fermi seas.
The single-particle energy is $e_i\equiv k_i^2\!/2m_i+U_i$,
and the indices $1,2$ collectively denote momentum,
isospin, and spin.
After the self-consistent BHF solution is obtained,
the energy density is evaluated as
\be
 \eps = \sum_{1<k_F^{(1)}}
 \qty( \frac{k^2}{2m_1} + \frac12 U_1(k) ) \:,
\label{eq:f}
\ee
and the nucleon chemical potentials follow
from the corresponding thermodynamic derivatives,
\be
 \mu_i = {\frac{\partial\eps}{\partial\rho_i}} \:.
\ee
The stellar matter is taken to be cold,
neutrino-free, charge neutral, and catalyzed.
Its composition contains neutrons, protons, electrons, and muons,
with the particle fractions fixed by weak-interaction $\beta$ equilibrium.
The pressure is then obtained
from the zero-temperature thermodynamic identity
\be
 p(\eps) = \rho^2 \frac{\partial}{\partial\rho}
 \frac{\eps(\rho_i(\rho))}{\rho}
 = \sum_i \rho_i \mu_i - \eps \:.
\ee

The microscopic input is specified by the nuclear interaction.
In this study,
we employ the Argonne $V_{18}$ (V18) potential \cite{Wiringa95}
supplemented by compatible microscopic three-body forces
\cite{Grange89,Zuo02,Li08a,Li08b}.
This combination gives a realistic saturation point
and associated bulk properties of symmetric NM.
For the numerical work,
the empirical energy-density parametrizations of Refs.~\cite{Lu19,Wei20}
are used as a convenient analytic representation of the underlying BHF results.

Since the BHF calculation adopted here applies to uniform core matter,
it must be matched to a low-density crust EOS in the nonuniform regime.
We attach the BHF core EOS to the Shen2020 crust EOS \cite{Shen20},
which provides the low-density nonuniform matter outside the uniform core.
The matching prescription is deliberately kept simple.
The transition is placed where the pressure and energy density
of the crust and core branches coincide.
Non-unified crust-core matching can affect the radii of low-mass NSs \cite{Zdunik16}.
For the stellar masses considered here, $M\geq1.0\ms$,
the dependence on the precise transition density is comparatively weak
\cite{Burgio10,Baldo14,Fortin16}.

\subsection{Dark matter}

We describe the DM component as a cold asymmetric fermion fluid.
Rather than choosing a unique dark-sector particle model,
a minimal EOS is adopted whose stiffness is set
by the fermion mass and a short-range repulsive self-interaction.
This choice is intended as a reference realization
rather than a unique description of the dark sector.
The dark particles are assumed to be stable on stellar time scales
and to couple to ordinary matter only gravitationally.
Consequently,
the ordinary and dark sectors have independent EOSs,
with no chemical equilibrium imposed between them.

Following the standard construction
for self-interacting fermionic stars
\cite{Narain06,Tulin13,Kouvaris15},
the microscopic input consists of a fermion mass $\mu$
and a Yukawa interaction
\be
 V_D(r)=\alpha_D\frac{e^{-m_Dr}}{r}\:,
\ee
where $\alpha_D$ is the dark coupling and $m_D$ is the mediator mass.
For uniform matter at zero temperature, define
\be
 x_D\equiv\frac{k_{F,D}}{\mu}
 =\frac{(3\pi^2n_D)^{1/3}}{\mu}\:,
\ee
where $n_D$ is the DM-particle number density.
The kinetic part is the relativistic degenerate Fermi-gas contribution,
and the repulsive interaction adds an identical positive mean-field term
to the pressure and energy density.
\bal
 p_D(n_D)
 &=\frac{\mu^4}{8\pi^2}
 \bigg[
 x_D\sqrt{1+x_D^2}\left(\frac{2x_D^2}{3}-1\right)
 +\operatorname{arsinh}x_D
 \bigg]
 +\delta_D\:,
\\
 \eps_D(n_D)
 &=\frac{\mu^4}{8\pi^2}
 \bigg[
 x_D\sqrt{1+x_D^2}(2x_D^2+1)
 -\operatorname{arsinh}x_D
 \bigg]
 +\delta_D\:.
\label{e:dm_eos}
\eal
It is convenient to absorb $\alpha_D/m_D^2$
into a dimensionless interaction parameter
\be
 y_D^2\equiv\frac{2\pi\alpha_D\mu^2}{m_D^2}\:,
\ee
so that
\bal
 \delta_D
 &=\frac{2}{9\pi^3}\frac{\alpha_D\mu^6}{m_D^2}x_D^6
 =\mu^4\left(\frac{y_D}{3\pi^2}\right)^2x_D^6
 =\left(\frac{y_Dn_D}{\mu}\right)^2\:.
\label{e:dm_delta}
\eal
The interaction term scales as $n_D^2$
and therefore stiffens the high-density DM EOS relative
to a free fermion gas.

For the reference model adopted below,
we follow Ref.~\cite{Wei26} and evaluate the transfer cross section
in the Born limit \cite{Tulin13,Kouvaris15,Maselli17},
\be
 \sigma_D=\frac{4\pi\alpha_D^2\mu^2}{m_D^4}
 =\frac{y_D^4}{\pi\mu^2}\:.
\ee
A representative self-scattering strength
is chosen in the range commonly considered for galaxy-cluster constraints
\cite{Markevitch04,Kaplinghat16,Sagunski21,Loeb22},
\be
 \frac{\sigma_D}{\mu}=1~\mathrm{cm}^2\,\mathrm{g}^{-1}
 =4560~\mathrm{GeV}^{-3}
\ee
fixes
\be
 y_D=\left(\pi\mu^3\frac{\sigma_D}{\mu}\right)^{1/4}
 \simeq10.94\left(\frac{\mu}{1~\mathrm{GeV}}\right)^{3/4}\:.
\ee
The numerical calculations below adopt
the representative value $\mu=1~\mathrm{GeV}$ used in Ref.~\cite{Wei26},
for which $y_D=10.94$.
The numerical EOS table is then generated
from Eqs.~(\ref{e:dm_eos}) and (\ref{e:dm_delta})
as a parametric function of $n_D$.
The adiabatic sound speed entering the perturbation equations
is $c_{s,D}^2=dp_D/d\eps_D$.
In the stellar models,
the amount of DM is specified macroscopically
by the target mass fraction $\fdm$,
rather than by an additional microscopic equilibrium condition
between the two fluids.

\section{Dark-Matter-Admixed Neutron Stars}
\label{s:osc}

\subsection{Hydrostatic configuration}
\label{s:tov}

In a DANS,
we model NM and DM as two perfect fluids
that interact only through the common spacetime metric.
The static,
spherically symmetric line element is
\be
 ds^2=-e^{\nu(r)}dt^2+e^{\la(r)}dr^2
 +r^2(d\theta^2+\sin^2\!\theta d\phi^2)\:,
\label{e:ds2}
\ee
where $e^{\nu(r)}$ and $e^{\la(r)}$ are metric functions.
Defining the total mass function and total pressure
by $m=m_\nm+m_\dm$ and $p=p_\nm+p_\dm$,
and setting $b\equiv m/r$ and $Q\equiv b+4\pi r^2p$,
the metric functions are determined by
\bal
 e^\la&=\frac{1}{1-2b}\:,
\\
 \frac{d\nu}{dr}&=\frac{2Q}{r(1-2b)}\:.
\label{e:metric}
\eal
Hydrostatic equilibrium is determined by the two-fluid TOV system
\cite{Kodama72,Comer99,Sandin09}
\bal
 \frac{dm_i}{dr}
 &=4\pi r^2\eps_i\:,
\\
 \frac{dp_i}{dr}
 &=-\frac{q_i Q}{r(1-2b)}\:.
\label{e:tov}
\eal
Here $i=\nm,\dm$ labels the NM and DM fluids,
$q_i\equiv p_i+\eps_i$, $\eps=\eps_\nm+\eps_\dm$,
and $b_i\equiv m_i/r$,
so that $b=b_\nm+b_\dm$.
The stellar radii $R_\nm$ and $R_\dm$
are defined by the vanishing of the respective pressures,
\be
 p_i(R_i)=0\:.
\label{e:tov_surfaces}
\ee
Equations~(\ref{e:tov}) are applied only where component $i$ is present.
For $r>R_i$, $p_i=\eps_i=0$ and $m_i$ is kept fixed at $M_i\equiv m_i(R_i)$.
The stellar radius is $R=\max(R_\nm,R_\dm)$,
while the visible radius is $R_\nm$.
The two possible geometries are DM-core stars ($R_\dm<R_\nm$)
and DM-halo stars ($R_\dm>R_\nm$).
The total gravitational mass is therefore
\be
 M=M_\nm+M_\dm=m_\nm(R_\nm)+m_\dm(R_\dm)\:.
\label{e:gm}
\ee
The DM mass fraction is defined as
\be
 \fdm=\frac{M_\dm}{M}\:.
\label{e:fdm}
\ee
To handle the two-fluid TOV problem,
we work with the relativistic enthalpy variable
of each component \cite{Lindblom10,Lindblom12},
defined by
\be
 h_i(p_i)=\int_0^{p_i}\frac{dp'}
 {p'+\eps_i(p')}\:.
\label{e:enthalpy}
\ee
Combining this definition with Eq.~(\ref{e:tov})
gives $dh_i/dr=-Q/[r(1-2b)]$ for every active component.
The same derivative appears because both fluids
feel the common gravitational potential generated
by the total pressure and total mass.
We therefore introduce a common enthalpy-drop coordinate $u$,
with $u=0$ at the center, such that
\bal
 h_i(u)&=h_{i,c}-u\:,
\\
 \frac{dr}{du}&=\frac{r(1-2b)}{Q}\:.
\label{e:u_relations}
\eal
These relations hold until the surface of fluid $i$ is reached.
When $u=h_{i,c}$, the pressure of component $i$ vanishes
and the corresponding surface $R_i$ is reached.
Since the two components share the same enthalpy-drop coordinate,
the formation of a DM core or a DM halo
is determined by the central enthalpies of the two components.
$h_{\dm,c}<h_{\nm,c}$ gives $R_\dm<R_\nm$,
whereas $h_{\dm,c}>h_{\nm,c}$ gives $R_\dm>R_\nm$.
If the two central enthalpies are equal,
the two surfaces coincide.
Thus the two component surfaces are reached
when the corresponding enthalpy vanishes
in the same integration coordinate.
For a prescribed value of $\fdm$,
we adjust the dark central enthalpy $h_{\dm,c}$
for each chosen ordinary central enthalpy $h_{\nm,c}$
until Eq.~(\ref{e:fdm}) is satisfied.
Varying $h_{\nm,c}$ then gives a stellar sequence
at fixed DM mass fraction.
These fixed-$\fdm$ curves provide the equilibrium backgrounds
used in the mode calculation.

For numerical integration,
we introduce the regular variables
$x\equiv r^2$ and $b_i=m_i/r$.
The equations are
\bal
 \frac{dx}{du}
 &=\frac{2x(1-2b)}{Q}\:,
\\
 \frac{db_i}{du}
 &=\frac{(4\pi x\eps_i-b_i)(1-2b)}{Q}\:.
\label{e:tov_u}
\eal
The center is initialized with the finite limits implied
by $u=A_c r^2+O(r^4)$,
where $A_c=2\pi(\eps_c+3p_c)/3$ and $\eps_c,p_c$ are total central values,
\be
 \left.\frac{dx}{du}\right|_c
 =\frac{3}{2\pi(\eps_c+3p_c)}\:,\qquad
 \left.\frac{db_i}{du}\right|_c
 =\frac{2\eps_{i,c}}{\eps_c+3p_c}\:.
\label{e:center_limits}
\ee
The metric potential $\nu$ follows from
\be
 \frac{d\nu}{du}
 =\frac{d\nu}{dr}\frac{dr}{du}=2\:.
\label{e:nu_u_derivative}
\ee
Its additive constant is fixed
by matching to the exterior Schwarzschild solution
at $u_R\equiv u(R)=\max(h_{\nm,c},h_{\dm,c})$,
\be
 \nu(u)=\ln\left(1-\frac{2M}{R}\right)+2(u-u_R)\:.
\label{e:nu_u}
\ee

\subsection{Non-radial oscillations}

\subsubsection{Perturbation equations}

Even-parity NROs are described in the Regge-Wheeler gauge.
The harmonic convention is $e^{i\om t}$,
so that an outgoing wave at infinity
is proportional to $e^{-i\om r_*}$ and
\be
 \om=2\pi f+i/\tau
\ee
with damping time $\tau$.
Thus $\om_R\equiv\Re\om=2\pi f$.
The metric perturbation is written as
\bal
 ds^2 =& -e^{\nu(r)} \big[
 1 + r^l H_0(r) e^{i\om t} Y^l_m(\theta,\phi) \big] dt^2
\non\\
       & +e^{\la(r)} \big[
 1 - r^l H_0(r) e^{i\om t} Y^l_m(\theta,\phi) \big] dr^2
\non\\
 & + \big[ 1 - r^l K(r) e^{i\om t} Y^l_m(\theta,\phi) \big]
 r^2 ( d\theta^2 + \sin^2\!\theta d\phi^2 )
\non\\
 & -2i\om r^{l+1} H_1(r) e^{i\om t}Y^l_m(\theta,\phi)\,dt\,dr\:.
\label{e:metric_pert}
\eal
The usual $H_2$ amplitude has been set equal to $H_0$,
as appropriate for perfect fluids without anisotropic stress.
Each fluid has its own Lagrangian displacement,
\bal
 \xi_i^r
 &=r^{l-1}e^{-\la/2}W_iY^l_m e^{i\om t}\:,
\non\\
 \xi_i^\theta
 &=-r^{l-2}V_i\partial_\theta Y^l_m e^{i\om t}\:,
\non\\
 \xi_i^\phi
 &=-r^{l-2}(\sin\theta)^{-2}V_i
 \partial_\phi Y^l_m e^{i\om t}\:,
\label{e:xi}
\eal
and its own Lagrangian-pressure variable,
\be
 \Delta p_i=-r^le^{-\nu/2}X_iY_m^le^{i\om t}\:.
\label{e:delta_p}
\ee

The single-fluid relativistic NRO equations are well established
in the Lindblom-Detweiler formulation
\cite{Thorne67,Lindblom83,Detweiler85,Zhao22b}.
For the DANS models considered here,
we follow the gravitationally
coupled two-fluid formulation given in Ref.~\cite{Kumar26}.
The two fluids are taken to be non-entrained
and to have no direct nongravitational coupling.
The evolved interior variables
are the common metric amplitudes $H_1$ and $K$,
together with the fluid variables $W_i$ and $X_i$
for each active component.
The remaining amplitudes $H_0$ and $V_i$ are fixed algebraically.
With $n\equiv(l-1)(l+2)/2$, the evolution equations are
\bal
 r\frac{dH_1}{dr}
 &=
 -\left[l+1+2be^\la+4\pi r^2e^\la(p-\eps)\right]H_1
\non\\
&\quad
 +e^\la\left(H_0+K-16\pi S_V\right)\:,
 \label{e:nro1}
\\
 r\frac{dK}{dr}
 &=
 H_0+(n+1)H_1
 +(e^\la Q-l-1)K
 -8\pi e^{\la/2}S_W\:,
 \label{e:nro2}
\\
 r\frac{dW_i}{dr}
 &=
 -(l+1)\left(W_i+l e^{\la/2}V_i\right)
\non\\
&\quad
 +r^2 e^{\la/2}
 \left[
 \frac{X_i}{q_ic_{s,i}^2e^{\nu/2}}
 +\frac{H_0}{2}+K
 \right]\:,
 \label{e:nro3}
\eal
\bal
 r\frac{dX_i}{dr}
 &=
 -lX_i
 +\frac{q_ie^{\nu/2}}{2}
 \Bigg\{
 (3e^\la Q-1)K
 -4(n+1)e^\la Q\frac{V_i}{r^2}
\non\\
&\quad
 +(1-e^\la Q)H_0
 +\left(\om^2r^2e^{-\nu}+n+1\right)H_1
\non\\
&\quad
 -\left[
 2\om^2e^{\la/2-\nu}
 -r^2\frac{d}{dr}
 \left(
 \frac{e^{-\la/2}}{r^2}
 \frac{d\nu}{dr}
 \right)
 \right]W_i
\non\\
&\quad
 -8\pi e^{\la/2}S_W
 \Bigg\}\:.
\label{e:nro4}
\eal
Here $c_{s,i}^2\equiv(\partial p_i/\partial\eps_i)_{\rm ad}$
is the adiabatic sound speed and the source terms are defined as
\be
 S_W\equiv\sum_{i}q_iW_i\:,\qquad
 S_V\equiv\sum_{i}q_iV_i\:.
\label{e:shortdefs}
\ee
The sums run over the fluid components
that are present at the radius under consideration.
The common term proportional to $S_W$ in the $X_i$ equation
is the explicit gravitational coupling
between the two fluids through the radial Einstein equation.
Since no direct microphysical coupling or entrainment is included,
all cross-component terms enter through the common metric source.

The variables $H_0$ and $V_i$ are not evolved independently.
At each radius they are obtained from the algebraic constraints
\bal
 H_0 &= (n+2b+Q)^{-1}\Big\{
 \big[(\om r)^2e^{-(\nu+\la)}-(n+1)Q\big]H_1
\non\\&\hskip2mm
 +\big[n-(\om r)^2e^{-\nu}-Q^2e^\la+Q\big]K
 +8\pi r^2e^{-\nu/2}\sum_{i}X_i
 \Big\}\:,
\\
 V_i &= \frac{e^\nu}{\om^2}
 \left[
 \frac{e^{-\nu/2}}{q_i}X_i
 -\frac{Q}{r^2}e^{\la/2}W_i
 -\frac12H_0
 \right]\:.
\label{e:alg}
\eal
Writing $Q=Q_2r^2+{\cal O}(r^4)$ near the center,
with $Q_2=4\pi(\eps_c+3p_c)/3$,
regularity of the perturbations fixes the leading coefficients
of a central solution as
\bal
 H_{1,c}
 &=\frac{lK_c+8\pi\sum_iq_{i,c}W_{i,c}}{n+1}\:,
\\
 X_{i,c}
 &=q_{i,c}e^{\nu_c/2}
 \left[
 \left(Q_2-\frac{\om^2}{l e^{\nu_c}}\right)W_{i,c}
 +\frac{K_c}{2}
 \right]\:.
\label{e:center}
\eal
At the surface of each component the Lagrangian pressure perturbation vanishes,
\be
 X_i(R_i)=0\:.
\label{e:surface}
\ee
If one component ends inside the other,
this condition is imposed at its own surface only.
The metric perturbations and the perturbations of the remaining fluid
are then continued to the outer stellar surface.

\subsubsection{Exterior matching}

For $r>R$,
the background is the Schwarzschild spacetime with mass $M$.
In this exterior region, $b=M/r$.
Introducing the tortoise coordinate $\rstar=r+2M\ln(r/2M-1)$,
the exterior even-parity perturbation
can be written in terms of the Zerilli function $Z$,
which satisfies \cite{Zerilli70}
\be
 \frac{d^2Z}{d\rstar^2}
 +\left[\om^2-V_Z(r)\right]Z=0\:,
\label{e:zerilli}
\ee
where the Zerilli potential is
\be
 V_{\rm Z}
 =2(1-2b)\frac{
 n^2(n+1)+3n^2b+9nb^2+9b^3
 }{(n+3b)^2r^2}\:.
\label{e:vz}
\ee
The metric amplitudes in Eq.~(\ref{e:metric_pert})
enter the physical metric
through the combinations $r^lK$ and $r^{l}H_1$.
The standard Zerilli
reconstruction gives these combinations as \cite{Lindblom83}
\bal
 r^lK
 &=g_{\rm Z}(r)\frac{Z}{r}+\frac{dZ}{d\rstar}\:,
 \label{e:zmetric1}
\\
 r^{l}H_1
 &=h_{\rm Z}(r)\frac{Z}{r}+(1-2b)^{-1}\frac{dZ}{d\rstar}\:,
 \label{e:zmetric2}
\\
 g_{\rm Z}(r)
 &=\frac{n(n+1)+3nb+6b^2}{n+3b}\:,
 \label{e:zmetric3}
\\
 h_{\rm Z}(r)
 &=\frac{n-3nb-3b^2}{(1-2b)(n+3b)}\:.
\label{e:zmetric4}
\eal
The exterior values of $K$ and $H_1$
are matched to the interior solution
at the outer stellar surface $R$.
The remaining exterior metric variable is fixed algebraically,
\bal
 H_0
 &=
 \frac{1}{(1-2b)(3b+n)}
 \Big\{
 \big[(\om r)^2-(n+1)b\big]H_1
\non\\
&\qquad
 +\big[n(1-2b)-(\om r)^2+b(1-3b)\big]K
 \Big\}\:.
\label{e:h0_ext}
\eal

At infinity, the outgoing Zerilli solution has the asymptotic form
\be
 Z(r)\rightarrow A_Z e^{-i\om\rstar},
 \qquad r\rightarrow\infty\:.
\label{e:azdef}
\ee
With the time dependence $e^{i\om t}$,
this corresponds to a wave proportional to $e^{i\om(t-\rstar)}$.
The coefficient $A_Z$ is the radiative Zerilli amplitude.
It is the exterior quantity entering the GW energy flux
and the strain estimate below.

Although the Zerilli function provides the radiative amplitude,
the outgoing-wave condition is imposed numerically
in the equivalent Regge-Wheeler representation,
following Refs.~\cite{Chandrasekhar75,Kumar26,Regge57}.
The corresponding master function satisfies
\be
 \frac{d^2\psi}{d\rstar^2}
 +\left[\om^2-V_{\rm RW}(r)\right]\psi=0\:,
\label{e:rw}
\ee
where the Regge-Wheeler potential is
\be
 V_{\rm RW}
 =(1-2b)\left[\frac{l(l+1)}{r^2}-\frac{6M}{r^3}\right]\:.
\label{e:vrw}
\ee
To connect the Regge-Wheeler solution
back to the exterior metric variables,
the Chandrasekhar transformation is applied
\cite{Chandrasekhar75,Kumar26},
\bal
 \big[4n(n+1)+12i\om M\big]Z
 &=C_{\rm Ch}\psi
 +12M\frac{d\psi}{d\rstar}\:,
\\
 \big[4n(n+1)-12i\om M\big]\psi
 &=C_{\rm Ch}Z
 -12M\frac{dZ}{d\rstar}\:,
\label{e:chandra}
\eal
where the radial coefficient is
\be
 C_{\rm Ch}
 =4n(n+1)+\frac{36M^2(r-2M)}{r^2(nr+3M)}\:.
\ee
With the convention $e^{i\om t}$,
an outgoing wave at infinity is proportional to $e^{-i\om\rstar}$.
Following the standard continued-fraction construction
\cite{Leaver90,Benhar99,Kumar26},
the RW solution is expanded about a finite vacuum radius $r_{\rm a}$,
\be
 \psi=e^{-i\om\rstar}
 \sum_{j=0}^{\infty}a_j\left(1-\frac{r_{\rm a}}{r}\right)^j\:.
\ee
Substitution into Eq.~(\ref{e:rw}) gives
\be
 \alpha_j\,a_{j+1}
 +\beta_j\,a_j
 +\gamma_j\,a_{j-1}
 +\delta_j\,a_{j-2}=0\:,
\label{e:rw_rec}
\ee
where $a_{-1}=0$, $j\ge1$, and
\bal
 \alpha_j&=\left(1-\frac{2M}{r_{\rm a}}\right)j(j+1)\:,
\\
 \beta_j&=-2\left[
 i\om r_{\rm a}+\left(1-\frac{3M}{r_{\rm a}}\right)j
 \right]j\:,
\\
 \gamma_j&=\left(1-\frac{6M}{r_{\rm a}}\right)j(j-1)
 +\frac{6M}{r_{\rm a}}-l(l+1)\:,
\\
 \delta_j&=\frac{2M}{r_{\rm a}}(j-3)(j+1)\:.
\eal
The first two coefficients are not independent for the minimal outgoing tail.
To obtain their ratio, Eq.~(\ref{e:rw_rec}) is first reduced
by Gaussian elimination to the three-term form
\be
 \hat\alpha_j\,a_{j+1}
 +\hat\beta_j\,a_j
 +\hat\gamma_j\,a_{j-1}=0\:.
\ee
The first row is unchanged, and for $j\ge2$,
\bal
 \hat\alpha_j&=\alpha_j\:,
\\
 \hat\beta_j&=\beta_j
 -\frac{\hat\alpha_{j-1}\delta_j}{\hat\gamma_{j-1}}\:,
\\
 \hat\gamma_j&=\gamma_j
 -\frac{\hat\beta_{j-1}\delta_j}{\hat\gamma_{j-1}}\:.
\eal
The continued fraction then fixes
\be
 {\cal R}_{\rm a}\equiv\frac{a_1}{a_0}
 =
 -\frac{\hat\gamma_1}{
 \hat\beta_1
 -\dfrac{\hat\alpha_1\hat\gamma_2}{
 \hat\beta_2
 -\dfrac{\hat\alpha_2\hat\gamma_3}{
 \hat\beta_3-\cdots}}}\:.
\ee
Since $d(1-r_{\rm a}/r)/dr=1/r_{\rm a}$ at $r=r_{\rm a}$,
this ratio gives the outgoing state at the expansion radius,
up to the arbitrary normalization $a_0$,
\bal
 \psi(r_{\rm a})&=e^{-i\om r_{*{\rm a}}}a_0\:,
\\
 \left.\frac{d\psi}{d\rstar}\right|_{r_{\rm a}}
 &=e^{-i\om r_{*{\rm a}}}a_0
 \left[
 -i\om+\left(1-\frac{2M}{r_{\rm a}}\right)
 \frac{{\cal R}_{\rm a}}{r_{\rm a}}
 \right]\:.
\eal
The normalization $a_0$ is arbitrary.
In the numerical implementation it is assigned a fixed nonzero value,
since any such choice can be absorbed
into the exterior matching amplitude.
The RW solution is then integrated inward
from $r_{\rm a}$to $R$ and mapped to $Z$ and $dZ/d\rstar$
by the Chandrasekhar transformation.
Eqs.~(\ref{e:zmetric1})-(\ref{e:zmetric4}) finally
determine the exterior metric amplitudes $K_{\rm ext}(R)$ and $H_{1,\rm ext}(R)$.
The arbitrary normalization of this exterior basis solution
is carried by a separate matching coefficient
and does not affect the determinant condition below.

\subsubsection{Numerical method and mode diagnostics}

The perturbation calculation starts
from the equilibrium background constructed
in Sec.~\ref{s:tov}.
The perturbation equations are integrated
with the same enthalpy-drop coordinate,
\be
 \frac{dY}{du}=\frac{dr}{du}\frac{dY}{dr},
 \qquad
 Y\in\{H_1,K,W_i,X_i\}\:.
\ee
Beyond a component surface,
that component no longer contributes
to the matter source terms
and its perturbation variables are no longer evolved,
while the other component continues
to the outer stellar surface.

At the center,
where both fluids are active,
there are three regular central amplitudes.
Besides the ordinary co-moving response,
the system also admits a relative counter-moving
degree of freedom between the NM and DM components.
We therefore choose a central basis
in which the third vector isolates a bounded counter-moving direction.
\be
\begin{array}{r@{\quad}r@{}l@{\qquad}r@{}l@{\qquad}r@{}l}
 \bm y_0:
 &K_c&=q_c,
 &W_{\nm,c}&=1,
 &W_{\dm,c}&=1\:,
\\
 \bm y_1:
 &K_c&=-q_c,
 &W_{\nm,c}&=1,
 &W_{\dm,c}&=1\:,
\\
 \bm y_2:
 &K_c&=q_c,
 &W_{\nm,c}&=\frac{q_{\dm,c}}{q_c},
 &W_{\dm,c}&=-\frac{q_{\nm,c}}{q_c}\:.
\end{array}
\label{e:basis}
\ee
where $q_c=q_{\nm,c}+q_{\dm,c}$.
The cross weighting in $\bm y_2$ makes it satisfy
$q_{\nm,c}W_{\nm,c}+q_{\dm,c}W_{\dm,c}=0$,
so that the leading center motion does not contribute
to the total source term $S_W$.
The integration starts at a small radius
determined by the next-to-leading-order analytic center expansion.

For each trial $\om$,
let $\bm y_k(r,\om)$,
with $k=0,1,2$,
denote the three center-regular solutions
integrated outward from the center to $R$.
The exterior outgoing state is generated
by the Regge-Wheeler continued-fraction construction described above.
An exterior radius $r_{\rm a}$ is chosen in the vacuum region,
where the finite-radius continued fraction
determines the outgoing logarithmic derivative.
Although the continued fraction
only requires $r_{\rm a}$ to be large enough
for the outgoing tail to be convergent,
$r_{\rm a}=50/\om_R$ is taken in geometrized units as a conservative choice.
This boundary condition is accepted
when the outgoing logarithmic derivatives
obtained with truncation orders $J=128$ and $J=96$
agree within the prescribed relative tolerance.
If this check fails,
a third-order RW asymptotic expansion
supplies the same outgoing logarithmic derivative.
Both procedures are numerical implementations
of the same exterior outgoing condition
and lead back to the Zerilli amplitude $A_Z$.
The RW equation is then integrated inward from $r_{\rm a}$
to $R$ and transformed to the exterior $K$ and $H_1$.

The physical interior eigenfunction
is the linear combination
$c_0\bm y_0+c_1\bm y_1+c_2\bm y_2$,
while $A_{\rm ext}$ multiplies the outgoing exterior solution.
Thus the matching system contains
three center-regular interior basis solutions
and one exterior outgoing basis solution.
Surface matching gives
\be
 {\cal M}(\om)
 \begin{pmatrix}
 c_0&c_1&c_2&A_{\rm ext}
 \end{pmatrix}^{T}=0\:,
\ee
where the rows impose the two component-surface conditions $X_i(R_i)=0$
and the continuity of $K$ and $H_1$ at $R$.
Non-trivial solutions exist when
\be
\det{\cal M}(\om)=0\:.
\label{e:det}
\ee
Complex roots of Eq.~(\ref{e:det})
are tracked along the stellar sequence,
and the associated null vector fixes the relative amplitudes
and the chosen eigenfunction normalization.
For very weakly damped modes,
$\om_I$ can approach the absolute tolerance of the complex root search.
The height and precise location of the narrowest damping-time peaks
are therefore treated as approximate numerical results.

The diagnostics introduced below
separate three features of a two-fluid eigenfunction.
The quantities $\eta_i$, $C_{ND}$, and $R_Q$ measure, respectively,
which component carries the fluid motion,
whether the two displacement fields are co-moving or counter-moving,
and whether the matter quadrupoles add or cancel.

The component content is first characterized
by the maximum kinetic energy carried by each fluid.
Extending the standard estimate
of the oscillation kinetic energy \cite{Thorne69b}
to the two non-entrained components,
we write
\be
 E_i=\frac12\om_R^2N_i\:,
\ee
where the component norm is
\be
 N_i=\int_0^{R_i}
 q_i e^{(\la-\nu)/2}r^{2l}
 \left(|W_i|^2+l(l+1)|V_i|^2\right)dr\:,
\ee
for a normal mode with oscillation frequency $\om_R$.
In the single-fluid problem,
this kinetic term is commonly used to
construct the total oscillation energy
by the equality of kinetic and potential contributions
over an oscillation cycle \cite{Thorne67}.
Here $E_i$ is retained only as a GR-weighted estimate
of the maximum oscillation kinetic energy carried by component $i$.
It is not the full canonical mode energy,
but it is adequate for comparing
the relative fluid-energy content of the two components.
The corresponding kinetic-energy fraction is defined as
\be
 \eta_i=\frac{E_i}{E_\nm+E_\dm}\:.
\ee
This fraction indicates whether a mode or sequence is NM-led, DM-led, or mixed.

To distinguish co-moving from counter-moving motion,
the relative phase of the two displacement fields
is evaluated only over the region occupied by both components,
with $R_{\min}\equiv\min(R_\nm,R_\dm)$.
The normalized overlap induced by the same bilinear form is
\be
\begin{aligned}
 C_{ND}
 =\frac{1}{\sqrt{N_\nm N_\dm}}\Re\int_0^{R_{\min}}
 &\sqrt{q_\nm q_\dm}\,
 e^{(\la-\nu)/2}r^{2l}
\\
 &\times
 \left(W_\nm^*W_\dm+l(l+1)V_\nm^*V_\dm\right)dr\:,
\end{aligned}
\label{e:cnd}
\ee
Positive (negative) $C_{ND}$ corresponds to
predominantly co-moving (counter-moving) fluid motion.

The matter quadrupole content is estimated
from the leading-order source mass quadrupole
used in the post-Newtonian (PN) quadrupole approximation
\cite{Thorne80,Szczepanczyk21},
\be
 Q_{ij}^{\rm PN}(t)
 =\int d^3x\,\delta\eps(t,{\bf x})
 \left(x_i x_j-\frac13\delta_{ij}r^2\right)\:.
\ee
Here $\delta\eps=\delta\eps_\nm+\delta\eps_\dm$
is the Eulerian energy-density perturbation.
For the quadrupole mode of a nonrotating star,
one may choose the $m=0$ representative.
Then $Q_{11}=Q_{22}=-Q_{33}/2$,
and the two component contributions are
\be
 Q_{33,i}^{\rm PN}
 =\frac{4}{3}\sqrt{\frac{\pi}{5}}
 \int_0^{R_i}\delta\eps_i(r)\,r^4dr\:.
\label{e:q33}
\ee
With the perturbation convention
$\delta\eps_i(t,{\bf x})=\delta\eps_i(r)Y_{20}e^{i\om t}$,
the equivalent $X_i,W_i$ form is obtained
from the linearized energy-conservation equation \cite{Thorne67}.
The radial amplitude entering Eq.~(\ref{e:q33}) is
\be
 \delta\eps_i
 =-\frac{r^2X_i}{c_{s,i}^2e^{\nu/2}}
 -re^{-\la/2}W_i\frac{d\eps_i}{dr}\:.
\ee
The total PN quadrupole is the sum of the two component contributions.
Its degree of cancellation is measured by
\be
 R_Q
 =\frac{|Q_{33,\nm}^{\rm PN}+Q_{33,\dm}^{\rm PN}|}
 {|Q_{33,\nm}^{\rm PN}|+|Q_{33,\dm}^{\rm PN}|}\:.
\label{e:quad_diag}
\ee
The individual component quadrupoles scale
with the arbitrary eigenfunction normalization,
whereas the ratio $R_Q$ is independent of this overall normalization.
The exterior GW radiation is determined
by the single outgoing vacuum metric perturbation,
while Eqs.~(\ref{e:cnd}) and (\ref{e:quad_diag}) are internal diagnostics
for classifying the two-fluid eigenfunctions.
For later reference,
we also define the relative phase of the two component quadrupoles by
\be
 C_Q
 =
 \frac{\Re\!\left(Q_{33,\nm}^{\rm PN}Q_{33,\dm}^{{\rm PN}*}\right)}
 {|Q_{33,\nm}^{\rm PN}|\,|Q_{33,\dm}^{\rm PN}|}\:.
\label{e:cq}
\ee
Thus $C_Q=-1$ corresponds to opposite quadrupole phases,
whereas $C_Q=+1$ corresponds to equal phases.
The quantity $C_Q$ is interpreted
together with the component quadrupole amplitudes,
since it is undefined
when one of the two component quadrupoles vanishes.

\subsubsection{Gravitational-wave emission and detectability}

Previous estimates of oscillation-induced GW strain
are often based on the PN mass quadrupole
constructed from the internal density perturbation
\cite{Zheng23,Zheng25a}.
In the present full-GR calculation,
this internal quadrupole is used
only as a diagnostic of the matter source.
The self-consistent radiative quantity
is instead the outgoing Zerilli amplitude $A_Z$ defined in Eq.~(\ref{e:azdef}).
This distinction is especially important in a two-fluid star,
where the exterior vacuum region contains
a single metric perturbation
rather than separate NM and DM wave amplitudes.

For a fixed eigenfunction normalization,
the period-averaged GW power carried by,
the outgoing Zerilli wave is \cite{Martel05}
\be
 P_{\rm GW}^{Z}
 =
 \frac{1}{128\pi}
 \frac{(l+2)!}{(l-2)!}
 \om_R^2 |A_Z|^2\:.
\label{e:pgw_z}
\ee
Here $\om_R\equiv\Re\om$ denotes
the real part of the complex eigenfrequency $\om$,
and is the oscillation frequency appearing in the wave phase.
The amplitude $A_Z$ scales with the arbitrary normalization
of the linear eigenfunction.
Therefore,
to quote an absolute strain,
we prescribe the total energy $E_{\rm GW}$
radiated by the mode in GWs.
Equivalently,
we assume that this energy is stored
in the excited mode
and is emitted through the GW channel during the ringdown
\cite{Flanagan98,Berti06}.
The remaining mode energy then decays as
\be
 E(t)=E_{\rm GW}e^{-2t/\tau}\:.
\label{e:emode_decay}
\ee
The corresponding cycle-averaged initial GW power is
\be
 P_{\rm GW}(0)
 =
 -\left.\frac{dE}{dt}\right|_{t=0}
 =
 \frac{2E_{\rm GW}}{\tau}\:.
\label{e:pgw_energy}
\ee
The same external master function
determines the waveform at large distance.
Using the even-parity waveform formula of Martel and Poisson \cite{Martel05},
the axisymmetric $m=0$ contribution
has vanishing cross polarization,
and its plus polarization is
\be
 h_+^{(l0)}(t,\theta)
 =
 \frac{A_Z e^{i\om_R t}}{2D}
 \left(
 \frac{\partial^2}{\partial\theta^2}
 -\cot\theta\frac{\partial}{\partial\theta}
 \right)Y_{l0}(\theta)\:.
\label{e:strain_az}
\ee
For $l=2$,
\be
Y_{20}(\theta) =
 \sqrt{\frac{5}{16\pi}}
(3\cos^2\theta-1),
\ee
and hence
\be
 \left(
 \frac{\partial^2}{\partial\theta^2}
-\cot\theta\frac{\partial}{\partial\theta}
 \right)Y_{20}(\theta)
= 6\sqrt{\frac{5}{16\pi}}\sin^2\theta .
\ee
Equation~(\ref{e:strain_az}) then gives
\be
 h_+^{(20)}(t,\theta)
 =
 \frac{3}{4}\sqrt{\frac{5}{\pi}}\,
 \frac{A_Z e^{i\om_R t}}{D}\sin^2\theta\:.
\label{e:hplus_theta}
\ee
The maximum plus-polarization amplitude
is reached at $\theta=\pi/2$,
which gives
\be
 h_{+,0}^{(20)}
 =
 \frac{3}{4}\sqrt{\frac{5}{\pi}}\,
 \frac{|A_Z|}{D}\:.
\label{e:hplus_az}
\ee
For $l=2$, Eq.~(\ref{e:pgw_z}) reduces to
\be
 P_{\rm GW}^{Z}
 =
 \frac{3}{16\pi}
 \om_R^2 |A_Z|^2\:.
\label{e:pgw_z_l2}
\ee
Equating this initial cycle-averaged power to Eq.~(\ref{e:pgw_energy})
determines the external amplitude,
\be
 |A_Z|
 =
 \left(
 \frac{32\pi E_{\rm GW}}
 {3\om_R^2\tau}
 \right)^{1/2}\:.
\label{e:az_energy}
\ee
Substituting Eq.~(\ref{e:az_energy}) into Eq.~(\ref{e:hplus_az})
removes the arbitrary eigenfunction normalization and gives
\be
 h_{+,0}^{(20)}
 =
 \frac{1}{\om_R D}
 \left(
 \frac{30E_{\rm GW}}{\tau}
 \right)^{1/2}\:.
\label{e:hplus_energy}
\ee
The corresponding waveform is modeled as
\be
h_+(t)
 =
 h_{+,0}^{(20)}
 e^{-t/\tau}\cos(\om_R t)\:,
\label{e:hplus_time}
\ee
up to angular and detector projection factors.
The numerical results adopt $E_{\rm GW}$ and $D$
as explicit parameters.
They are converted to geometrized units
when inserted in Eq.~(\ref{e:hplus_energy}).

The detector strain is
\be
 h_{\rm det}(t)
 =
 F_+h_+(t)+F_\times h_\times(t)\:.
\ee
For the $m=0$ convention used here,
we quote the maximum-orientation value by default.
An angle-averaged estimate can be obtained by replacing $h_{+,0}^{(20)}$
with $\kappa_{\rm ang}h_{+,0}^{(20)}$.
For random inclination, sky position,
and polarization angle in a single L-shaped interferometer,
the RMS factor for this $m=0$ convention is $\kappa_{\rm ang}=\sqrt{8/75}$
\cite{Finn93,Jaranowski98,Sathyaprakash09,Schutz11}.

Following Refs.~\cite{Finn92,Iacovelli22a},
the matched-filter signal-to-noise ratio (SNR) for detector $i$ is
\be
 {\rm SNR}_i^2
 =
 4\int_0^\infty
 \frac{|\tilde h_{{\rm det},i}(f)|^2}{S_{n,i}(f)}df\:,
\label{e:snr_def}
\ee
where $i$ labels the detector,
and $S_{n,i}(f)$ is its one-sided noise power spectral density.
For a narrow damped sinusoid with $f=\om_R/(2\pi)$,
the standard narrow-band ringdown approximation \cite{Flanagan98,Berti06} gives
\be
 {\rm SNR}_i
 \simeq
 \frac{\kappa_{\rm ang}h_{+,0}^{(20)}}{\sqrt{S_{n,i}(f)}}
 \left[
 \frac{\tau}{2}
 \left(1-e^{-2T_{\rm coh}/\tau}\right)
 \right]^{1/2}\:.
\label{e:snr_narrow}
\ee
Substituting Eq.~(\ref{e:hplus_energy}) into Eq.~(\ref{e:snr_narrow})
gives the SNR directly in terms of the radiated energy and source distance,
\be
 {\rm SNR}_i^2
 =
 \frac{15\kappa_{\rm ang}^2 E_{\rm GW}}
 {\om_R^2 D^2 S_{n,i}(f)}
 \left(1-e^{-2T_{\rm coh}/\tau}\right)\:.
\label{e:snr_energy}
\ee
Here $E_{\rm GW}$ and $D$ are input parameters,
whereas $\om_R$ and $\tau$ are obtained from the mode calculation.
For a prescribed threshold ${\rm SNR}_{\rm th}$,
the required radiated energy is therefore
\be
 E_{\rm GW}^{\rm req}
 =
 \frac{{\rm SNR}_{\rm th}^2\om_R^2D^2S_{n,i}(f)}
 {15\kappa_{\rm ang}^2
 \left(1-e^{-2T_{\rm coh}/\tau}\right)}\:.
\label{e:energy_required}
\ee
Here $T_{\rm coh}$ is the time over
which the signal can be coherently tracked.
If $T_{\rm coh}\gg\tau$,
the effective integration time approaches $\tau/2$.
If $T_{\rm coh}\ll\tau$,
it approaches $T_{\rm coh}$.
Thus,
for modes with $\tau\gg T_{\rm coh}$,
the coherent integration time,
rather than the damping time,
sets the accumulated SNR.
In the numerical estimates,
the adopted value of $T_{\rm coh}$ is specified
together with the detector sensitivity curves.
The fully coherent limit is recovered by taking $T_{\rm coh}\to\infty$.

\section{Numerical results}
\label{s:res}
In this section we present the equilibrium sequences
and the quadrupolar $f$-like modes
obtained from the full two-fluid calculation.
Unless otherwise stated,
the DM component is described by the fermionic EOS specified in Sec.~\ref{s:eos},
with $\mu=1~{\rm GeV}$ and $y_D=10.94$.
In the sequence figures below,
colors label the fixed DM mass fraction $\fdm=M_{\dm}/M$.
We organize the results in fixed-$\fdm$ sequences.
Along each sequence,
the ordinary-matter central enthalpy
is varied while $\fdm$ is held fixed by adjusting the DM central enthalpy.
Comparing these sequences helps distinguish evolution
along an equilibrium sequence from changes
associated with the prescribed global DM fraction.
The plotted fixed-fraction sequences sample the transition
from the ordinary-star limit to increasingly DM-rich configurations.
Accordingly,
terms such as small, moderate, and large DM fraction
are used as shorthand for the ordered set of fixed-$\fdm$ sequences
shown in the figures.
The two principal $f$-like continuations
are denoted $\fgl$ and $\frel$ below.
In a two-fluid star,
the ordinary-matter and DM displacement fields
have separate node structures,
so neither label asserts that both component eigenfunctions
remain node-free along the whole sequence.
Rather,
the labels identify the two continuations
followed from the $f$-mode sector of the spectrum;
their local component content and node structure
are diagnosed separately.
The $\fgl$ branch is connected to the ordinary single-fluid $f$ mode
in the limit $\fdm\to0$,
whereas $\frel$ denotes the additional two-fluid $f$-like branch
associated with the relative degree of freedom.
The one-fluid limits provide a consistency check on the branch continuation.
At $\fdm=0$, the $\fgl$ sequence reproduces
the full-GR $f$ mode of the ordinary V18 star,
whereas at $\fdm=1$ it becomes
the corresponding $f$-mode sequence of pure DM stars.
The $\frel$ continuation requires
both fluids to be present
and therefore has no one-fluid limit.
The symbols $\fgl$ and $\frel$ are therefore branch labels,
not fixed statements about the local NM/DM content
or the radiative efficiency of every point.
We first discuss the equilibrium structure,
then the real frequencies and internal mode character,
and finally the full-GR damping times and detector-oriented energy estimates.

\subsection{Equations of state and equilibrium sequences}

\begin{figure}[t]
\centering
\includegraphics[width=\columnwidth]{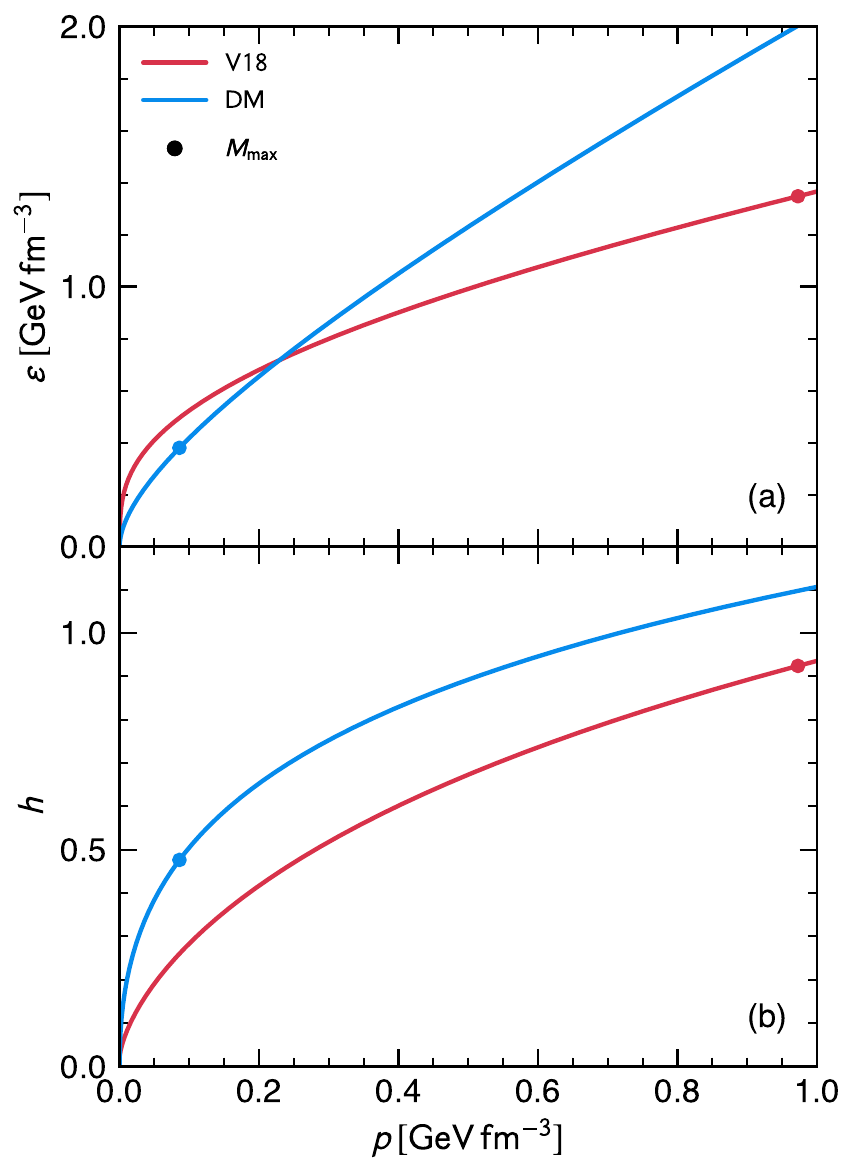}
\caption{
Panels (a) and (b) show, respectively, the energy density $\eps$
and relativistic enthalpy $h$ as functions of pressure for the
ordinary V18 EOS and for the DM fermion EOS used in this work.
The filled markers indicate
the central values of the maximum-mass single-component configurations.
}
\label{f:tf_eos}
\end{figure}

Figure~\ref{f:tf_eos} compares the ordinary-matter
and DM EOSs in the forms used in the two-fluid calculation.
The two curves are independent component EOSs,
rather than branches of a single effective or mixed EOS.
We evaluate $p_{\nm}(\eps_{\nm})$ and $p_{\dm}(\eps_{\dm})$
separately in the equilibrium and perturbation equations.
Their crossing in panel (a)
reflects the different density dependences
of the nuclear interaction and of the DM Fermi
and self-interaction contributions.
It should not by itself be interpreted as a reversal of the relative stiffness,
which is instead set by the local sound speeds $dp_i/d\eps_i$.

Panel (b) shows the component enthalpy-pressure relations used
to construct the backgrounds. During the integration, each EOS maps
its own enthalpy to pressure and energy density, while the two
components share the same enthalpy-drop coordinate. The relative
central enthalpies determine which component reaches zero pressure
first, thereby producing either DM-core or DM-halo configurations.

\begin{figure}[t]
\centering
\includegraphics[width=\columnwidth]{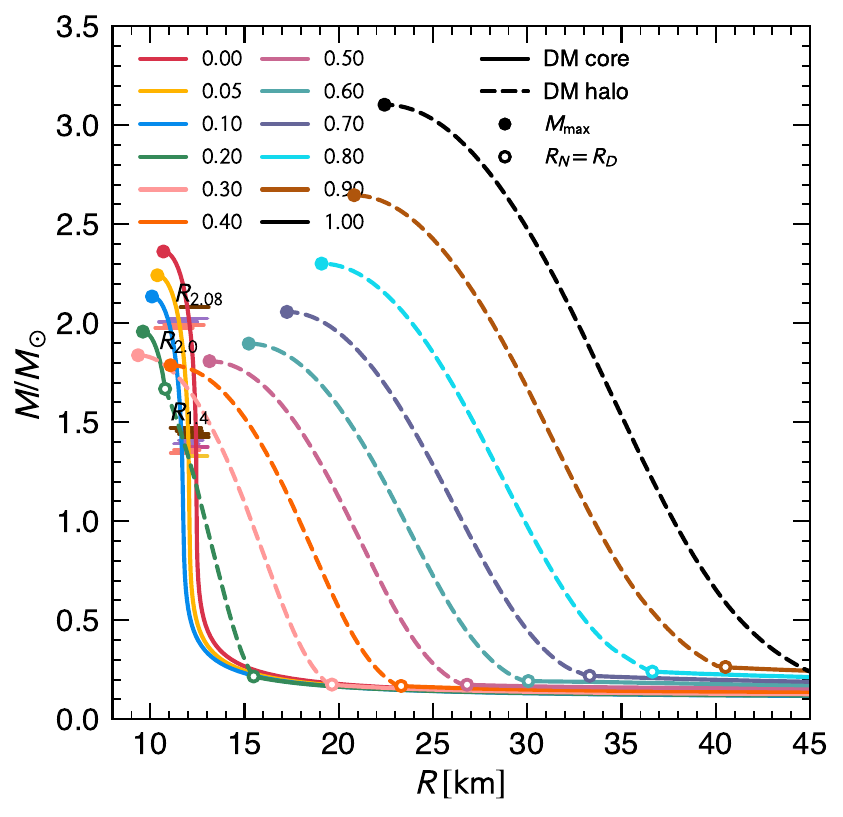}
\caption{
Mass-radius relations for fixed DM mass fractions $\fdm=M_{\dm}/M$.
The legend gives the corresponding values of $\fdm$,
and the radius is the outer radius $R=\max(R_{\nm},R_{\dm})$.
Filled circles mark the maximum-mass configurations along each sequence.
The horizontal bars indicate the limits
on $R_{2.08}$, $R_{2.0}$, and $R_{1.4}$
obtained from combined NICER and GW170817 analyses
\cite{Miller21,Pang21,Raaijmakers21,Rutherford24,Dittmann24,Mauviard25,Miller26}.
}
\label{f:tf_mr}
\end{figure}

The corresponding mass-radius curves
are shown in Fig.~\ref{f:tf_mr}.
For a two-fluid star,
the radial-stability boundary is formally defined
by the vanishing of the fundamental radial-mode frequency
under independent conservation
of the two component particle numbers.
For gravity-only two-fluid configurations,
radial-mode calculations show that this boundary
lies very close to the maximum of $M$ along fixed-$\fdm$ sequences \cite{Wei26}.
We therefore use the maximum-mass points
marked in Fig.~\ref{f:tf_mr} as practical indicators of the onset of radial instability.
The horizontal bars show the tightened radius intervals inferred
from combined analyses of NICER mass-radius measurements together with GW170817
\cite{Miller21,Pang21,Raaijmakers21,Rutherford24,Dittmann24,Mauviard25,Miller26}.
For small DM fractions the sequences remain
close to the ordinary V18 sequence in radius,
while their maximum masses decrease.
In this regime the DM fluid forms a compact component
inside the ordinary fluid,
and the additional central mass deepens the gravitational potential.
At larger $\fdm$ the outer radius grows rapidly
and the sequence enters a DM-halo regime.
The pure DM-star limit then reaches a much larger radius
and a larger maximum mass than the ordinary sequence.
For the present V18 ordinary EOS and reference DM EOS,
the maximum mass decreases from $M_{\rm max}=2.36\ms$ at $\fdm=0$
to about $1.79\ms$ around $\fdm=0.4$,
and then rises to $3.10\ms$ in the pure DM-star limit.
The radius at the maximum-mass point changes
from $R\simeq10.7\km$ in the ordinary limit to
$R\simeq22.4\km$ for the pure DM sequence.
Near the maximum-mass configurations,
the transition from a DM core to a DM halo
occurs between $\fdm=0.2$ and $\fdm=0.3$.
This nonmonotonic behavior is important
for interpreting the mode spectrum,
because the same total mass can correspond
to very different component radii,
average densities,
and relative fluid inertias.
Consequently one should not expect the two-fluid oscillation sequences
to follow the same ordering as a sequence of ordinary one-fluid stars.

\begin{figure}[t]
\centering
\includegraphics[width=\columnwidth]{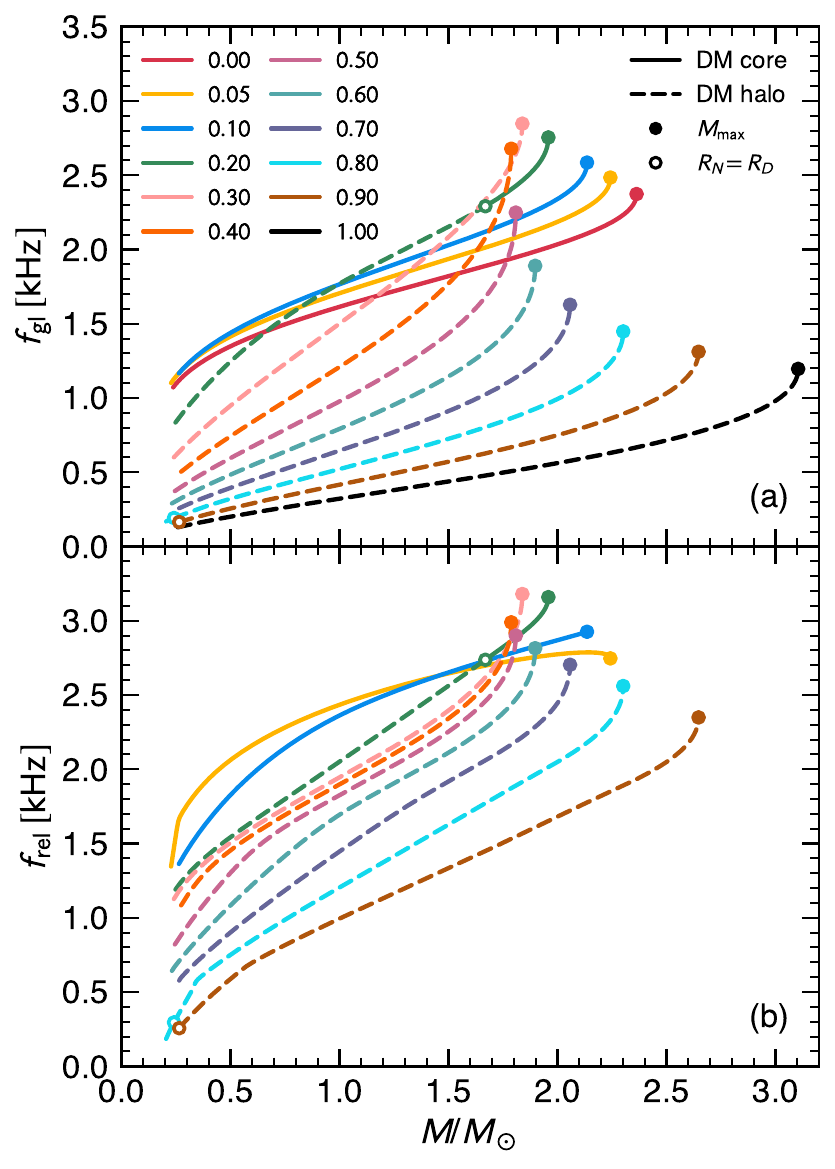}
\caption{
Real frequencies of the two main quadrupolar $f$-like sequences
as functions of the gravitational mass.
Panels (a) and (b) show the $\fgl$ and $\frel$ branches, respectively.
Colors label the fixed DM mass fraction $\fdm$,
with the legend giving the corresponding values.
Filled circles indicate the maximum-mass configurations.
}
\label{f:tf_freq}
\end{figure}

\begin{figure*}[t]
\centering
\includegraphics[width=0.83\textwidth]{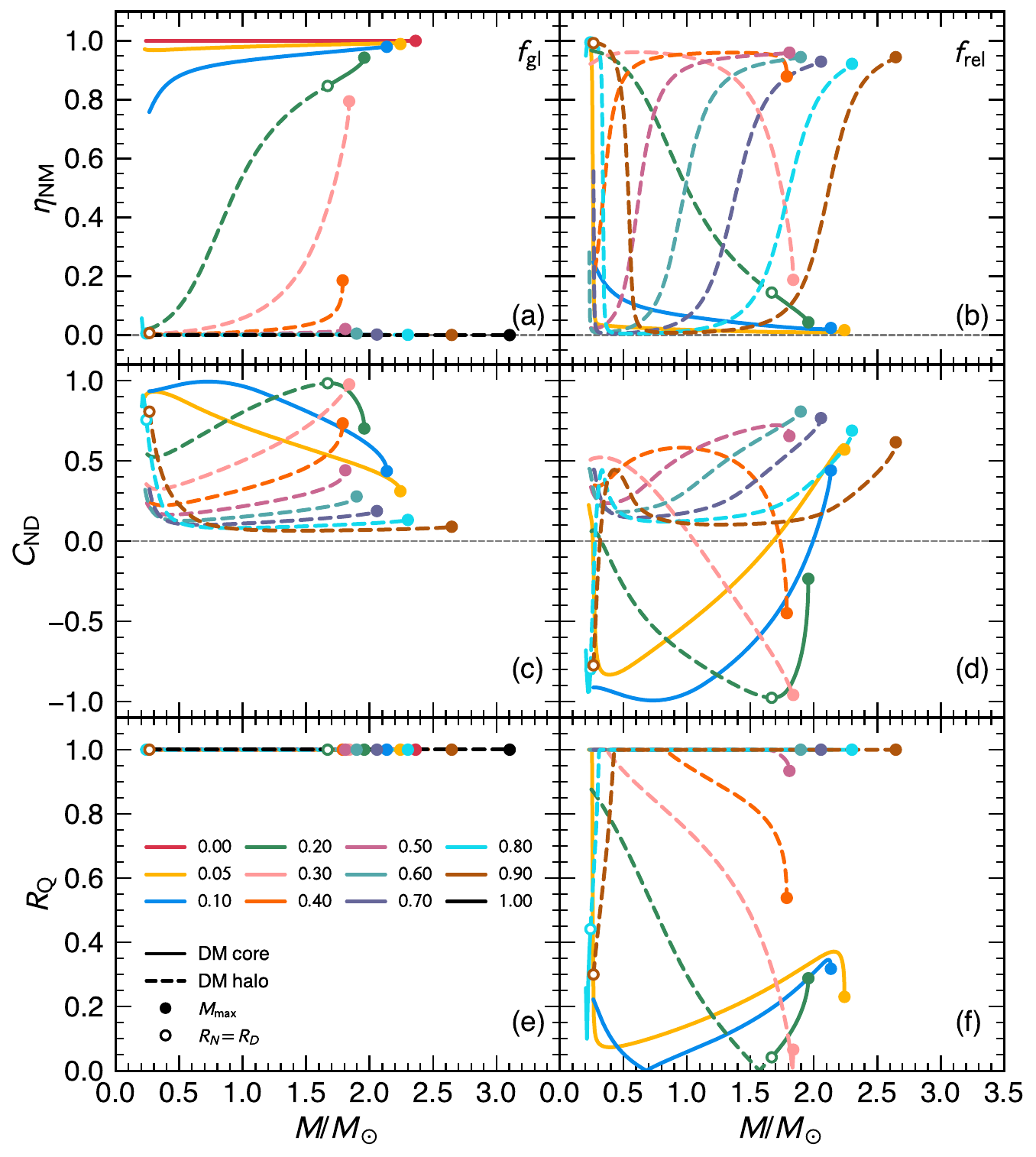}
\caption{
Mode diagnostics for the two principal $f$-like sequences.
Panels (a) and (b) show the ordinary-matter kinetic-energy fraction
$\eta_{\nm}=E_{\nm}/(E_{\nm}+E_{\dm})$.
Panels (c) and (d) show the normalized displacement overlap $C_{ND}$,
where positive and negative values indicate
predominantly co-moving and counter-moving motion.
Panels (e) and (f) show the matter-quadrupole cancellation ratio $R_Q$.
The left column, panels (a), (c), and (e), corresponds to $\fgl$,
and the right column, panels (b), (d), and (f), to $\frel$.
Colors label the fixed DM mass fraction $\fdm$,
with the legend giving the corresponding values.
Filled circles mark the maximum-mass configurations.
}
\label{f:tf_diag}
\end{figure*}

\begin{figure*}[t]
\centering
\includegraphics[width=0.83\textwidth]{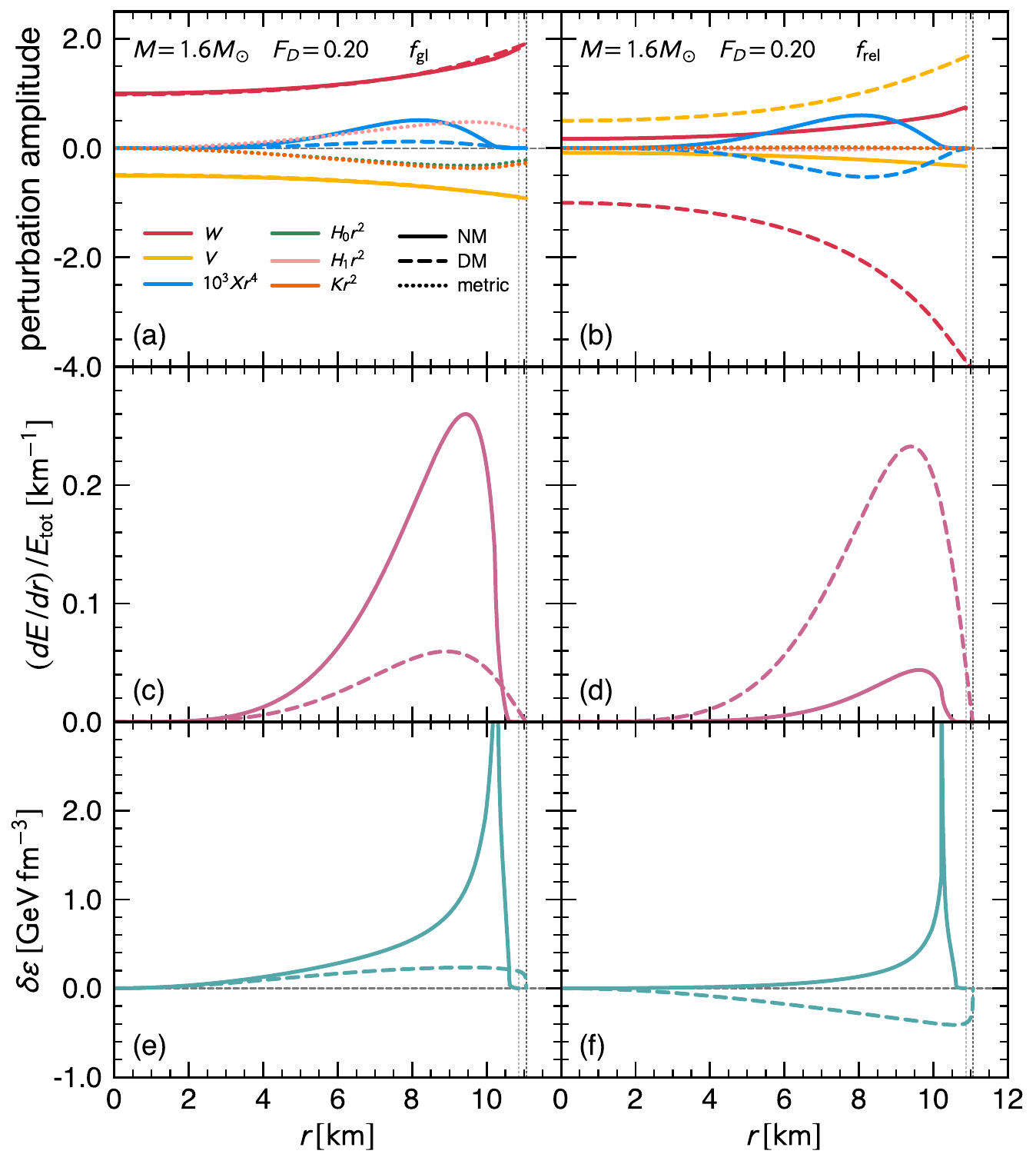}
\caption{
Representative radial profiles for the two $f$-like modes
of the same background model with $M=1.6\,M_\odot$ and $\fdm=0.20$.
Panels (a), (c), and (e) show $\fgl$, whereas panels (b), (d), and (f)
show $\frel$. Panels (a) and (b) show the real parts of the normalized
perturbation amplitudes
$W_i$, $V_i$, $10^3X_ir^4$,
and the common metric perturbations $H_0r^2$, $H_1r^2$, and $Kr^2$.
Solid and dashed curves denote ordinary-matter and DM fluid variables,
whereas dot-dashed curves denote metric variables.
The eigenfunctions are normalized
so that the larger central value of $|W_{\nm}|$ and $|W_{\dm}|$ is unity.
For the model shown here,
this sets $|W_{\nm}(0)|=1$ in the $\fgl$ column
and $|W_{\dm}(0)|=1$ in the $\frel$ column;
the displayed signs follow the chosen eigenfunction phase.
Panels (c) and (d) show the component kinetic-energy profiles $(dE_i/dr)/E_{\rm tot}$,
where $E_{\rm tot}=E_{\nm}+E_{\dm}$.
Panels (e) and (f) show the Eulerian energy-density perturbations
$\delta\epsilon_i$ in ${\rm GeV\,fm^{-3}}$
with the same eigenfunction normalization.
The vertical dotted lines mark the ordinary-matter and DM surfaces.
For this model the smaller radius is
$R_{\nm}$ and the larger one is $R_{\dm}$.
}
\label{f:tf_profile}
\end{figure*}

\subsection{Two principal \texorpdfstring{$f$}{f}-like sequences}

Figure~\ref{f:tf_freq} displays the real frequencies $f=\om_R/(2\pi)$
of the two principal $f$-like sequences.
Only these two principal $f$-like continuations
are shown in the main figures.
Other roots tracked by the numerical procedure
are outside the scope of the present discussion.
The one-fluid limits $\fdm=0$ and $\fdm=1$
are shown only for $\fgl$,
because the relative branch requires both components to be present.
In panel (a), the $\fgl$ branch is continuously connected
to the ordinary single-fluid $f$ mode when $\fdm\to0$.
At small and moderate $\fdm$ it remains the lower-frequency,
more global branch over most of the stable mass range.
Its frequency generally increases with mass,
following the usual trend with increasing mean density,
but the curve is displaced as the DM component modifies
the total radius and the gravitational potential.
The ordinary limit spans roughly $1.1$--$2.4\khz$
over the plotted mass range.
At larger DM fraction,
the same continuation increasingly follows the extended DM component,
and its frequency scale can drop substantially.
In the pure DM-star limit the plotted branch covers only about $0.13$--$1.20\khz$,
reflecting the much larger stellar radius.

Panel (b) shows the $\frel$ branch generated by the two-fluid problem.
It has no ordinary single-fluid counterpart
in the limit where the DM fluid is absent.
At small DM fraction it appears at a higher frequency
than $\fgl$ over much of the sequence,
which is consistent with a mode controlled
by a shorter component length scale or by the relative motion
between the two fluids.
At larger $\fdm$ the ordering becomes more model dependent.
The extended DM halo can drive a low-frequency DM-led response,
whereas the ordinary component can support a more compact,
higher-frequency motion.
For $\fdm=0.05$--$0.3$,
the upper parts of the $\frel$ curves reach $f\simeq2.7$--$3.2\khz$.
This higher scale is not simply a shift of the ordinary $f$ mode.
It reflects the additional relative degree of freedom,
whose effective dynamical length can be shorter than the radius of the whole star.

A useful qualitative comparison can nevertheless
be made with the two-fluid Cowling results.
In a DM-core configuration,
the lower-frequency $\fgl$ continuation is naturally
associated with the outer ordinary-matter Cowling $f$ mode,
whereas the higher-frequency $\frel$ continuation
is associated with the inner DM branch.
When the background becomes a DM halo,
the outer and inner roles are interchanged.
The $\fgl$ continuation is then naturally
compared with the DM-associated Cowling branch,
whereas $\frel$ is compared
with the ordinary-matter-associated one \cite{Sotani25a,Sotani25b}.

In the Cowling approximation,
the metric perturbations are neglected,
so these component-associated oscillations
are independent at linear order.
In full GR,
the two displacement fields are coupled
through the perturbed spacetime
and share a common complex eigenfrequency \cite{Kumar26}.
The correspondence above concerns
the continuation and spatial support of the branches,
rather than a one-to-one equality of their frequencies
or component amplitudes.
The Cowling solutions do not define a relative phase between the two fluids,
whereas the co-moving or counter-moving character
of a full-GR mode is fixed only
after the metric-mediated coupling and surface matching are imposed.

The frequency splitting is therefore only
the first level of the two-fluid information.
The same real-frequency branch
can change its component content along the sequence,
and two branches with comparable frequencies
may have very different radiative efficiencies.
For this reason the branch labels are used
as compact continuation labels.
The physical character is assigned from the component energy content,
the displacement overlap,
the quadrupole-cancellation diagnostic,
and the full-GR damping time discussed below.

\subsection{Mode diagnostics and quadrupole cancellation}

The diagnostic quantities in Fig.~\ref{f:tf_diag}
show how the two-fluid mode character changes along the fixed-$\fdm$ sequences.
The $\fgl$ diagnostics in panels (a), (c), and (e) show that at small
and moderate $\fdm$,
the ordinary-matter energy fraction is close to unity
over a large part of the sequence.
The displacement overlap is mostly positive,
and the quadrupole ratio remains near $R_Q\simeq1$.
In this parameter range,
this identifies $\fgl$ as an ordinary-matter-led,
predominantly co-moving continuation of the standard $f$ mode.
The quadrupole ratio remains essentially unity, $R_Q\simeq1$.
For all mixed $\fgl$ sequences considered here,
the component quadrupoles have the same phase,
$C_Q\simeq1$, and therefore add constructively.
One component can be subdominant at small or large $\fdm$,
but where both contributions are appreciable
they do not produce matter-quadrupole cancellation.
At larger DM fraction the same continuation
becomes DM-led rather than ordinary-led.
For example, the high-$\fdm$ curves have
$\eta_{\nm}\approx 1$ over most of their length,
while $C_{ND}$ remains positive
and $R_Q$ stays close to unity.
Thus the $\fgl$ label denotes
a global-like continuation,
not necessarily an NM-dominated or strictly additive branch.

The $\frel$ branch, shown in panels (b), (d), and (f), behaves differently.
Its energy content can be DM-led, NM-led,
or mixed depending on the mass and the DM fraction.
The overlap $C_{ND}$ often becomes negative,
showing that the two displacement fields are predominantly counter-moving.
However,
panels (e) and (f) demonstrate that counter-moving motion
alone is not enough to guarantee weak GW emission.
The relevant source-level condition is instead a small value of $R_Q$,
which requires the ordinary-matter and DM quadrupoles
to be nearly opposite in phase and comparable in magnitude.
Where this happens,
the total matter quadrupole is strongly suppressed
even though the individual component motions are not small.
These regions are therefore natural candidates for a weakly radiating $f$-like segment.
The strongest source-level cancellations in the present data occur
for intermediate DM fractions.
For $\fdm=0.1$, $0.2$, and $0.3$,
the minimum values of $R_Q$ on $\frel$
are of order $4\times10^{-4}$, $6\times10^{-4}$,
and $8\times10^{-4}$, respectively.
At these points $C_{ND}\simeq-1$ and $|Q_{\dm}/Q_{\nm}|\simeq1$,
consistent with two nearly opposite matter-quadrupole sources.
For $\fdm\gtrsim0.5$ this simple cancellation is much less prominent.
The branch can remain mixed,
or even mostly co-moving, and $R_Q$ often stays close to unity.

The two diagnostics separate kinematics from radiation source structure.
$C_{ND}$ follows the relative motion of the displacement fields,
while $R_Q$ follows the cancellation of the Eulerian matter-quadrupole source.
A counter-moving mode can still radiate efficiently
if one component dominates the quadrupole.
Likewise,
a mode with only moderate displacement anticorrelation
can radiate weakly when the density profiles
and radial eigenfunctions make the integrated quadrupoles cancel.
For this reason,
$\fgl$ and $\frel$ are used as continuation labels
rather than fixed physical mode names.
At low DM fraction,
$\fgl$ is naturally interpreted
as the ordinary-like global $f$ mode,
whereas $\frel$ often carries the additional relative two-fluid degree of freedom.
At larger DM fractions,
or near the transition between DM-core and DM-halo structures,
the character can change along the same continuation.
A branch can move from NM-led to mixed or DM-led,
and the sign and magnitude of the displacement overlap can vary.
The local terms ``NM-led'', ``DM-led'', ``co-moving dominated'',
``counter-moving dominated'', and ``quadrupole-cancelling''
are therefore more informative than
assigning a permanent component name to a whole branch.

The representative profiles in Fig.~\ref{f:tf_profile}
show how these integrated diagnostics arise from the eigenfunctions.
The two columns use the same equilibrium model,
so the difference is only the mode eigenfunction.
The vertical lines show that this representative model has $R_{\nm}<R_{\dm}$,
so the ordinary component terminates inside the DM halo.
Outside $R_{\nm}$ only the DM fluid variables remain,
whereas the metric perturbation is still
the single common spacetime response
sourced by both components in the interior.

For $\fgl$ [panels (a), (c), and (e)],
the ordinary-matter and DM displacement variables
have a largely compatible phase.
The kinetic-energy density is concentrated mainly
in the ordinary component,
and the Eulerian density perturbations of the two components
have the same sign over most of the overlap region.
The common metric perturbations are also visibly
larger than in the relative branch.
This is the profile-level counterpart
of the absence of strong matter-quadrupole cancellation
indicated by $R_Q\simeq1$.

For $\frel$ [panels (b), (d), and (f)], the two fluids move
in a more relative pattern.
The kinetic-energy density is mostly carried
by the DM component in this model,
but the ordinary component still contributes
appreciably to the density perturbation.
The two Eulerian density perturbations
have opposite signs over much of the common interior.
At the same time,
the dot-dashed metric curves in panel (b)
are strongly suppressed for the same eigenfunction normalization.
Thus the mode is not weak because the fluid motion is small.
Rather,
the two component source profiles drive a much weaker common metric perturbation.

The profiles provide a two-fluid example
of a source-cancellation mechanism also encountered
in nonradial oscillations of one-fluid stars.
In one-fluid $g$-mode calculations,
sign changes of the Eulerian density perturbation
can make the quadrupole source integral small
even when the local perturbation is not small,
leading to very long GW damping times
\cite{Zheng23}.
Here the cancellation is not only radial,
but also compositional:
the ordinary-matter and DM source profiles
can interfere destructively in the matter-quadrupole channel.
The sharp surface features in $\delta\epsilon_i$
should therefore be read as part of the source profile
entering the integral diagnostics,
not as evidence that the mode amplitude is small.
Figure~\ref{f:tf_profile} illustrates this source-level mechanism for one selected point.
Other long-lived configurations need not be associated
with source-level matter-quadrupole cancellation.
Their interpretation requires comparing the interior metric perturbations,
the surface combination that matches onto the Zerilli function,
and the outgoing amplitude $A_Z$.

\begin{table*}[t]
\caption{
Diagnostics of the two prominent damping-time peaks
on the $\frel$, $\fdm=0.10$ sequence.
The low-mass peak is associated
with source-level matter-quadrupole cancellation,
whereas the high-mass peak illustrates a surface radiative-projection minimum.
The amplitude columns use the same eigenfunction normalization
as the radial profiles.
The quantity
$\mathcal{R}_Z=\om_R^2|A_Z|^2/(E_{\nm}+E_{\dm})$
is normalization independent and is proportional
to the Zerilli-wave luminosity per unit of the adopted mode-energy estimate.
Here $R_{\rm in}$ is the inner component surface;
for these two models it is the DM surface,
while $R$ is the outer ordinary-matter surface.
The quoted damping times are approximate
because the corresponding imaginary frequencies
approach the numerical accuracy of the root search.
}
\renewcommand{\arraystretch}{1.2}
\begin{ruledtabular}
\begin{tabular}{c|cccccc|ccc}
Case
& $M/\ms$
& $\tau$ (s)
& $R_Q$
& $C_Q$
& $|A_Z|$
& $\mathcal{R}_Z$
& $|H_1r^2|_{\max}$
& $|H_1r^2|_{R_{\rm in}}$
& $|H_1r^2|_R$ \\
\hline
source-level
& $0.708$
& $\simeq2.8\times10^{7}$
& $6.09\times10^{-3}$
& $-1.00$
& $3.94$
& $8.49\times10^{-43}$
& $2.98\times10^{-3}$
& $6.47\times10^{-5}$
& $4.42\times10^{-5}$ \\
surface-projection
& $2.134$
& $\simeq6.5\times10^{7}$
& $3.31\times10^{-1}$
& $-1.00$
& $3.84\times10^{-1}$
& $1.09\times10^{-45}$
& $1.60\times10^{-1}$
& $1.15\times10^{-1}$
& $7.55\times10^{-6}$ \\
\end{tabular}
\end{ruledtabular}
\label{t:weak_diag}
\end{table*}

\subsection{Full-GR damping times}

\begin{figure}[t]
\centering
\includegraphics[width=\columnwidth]{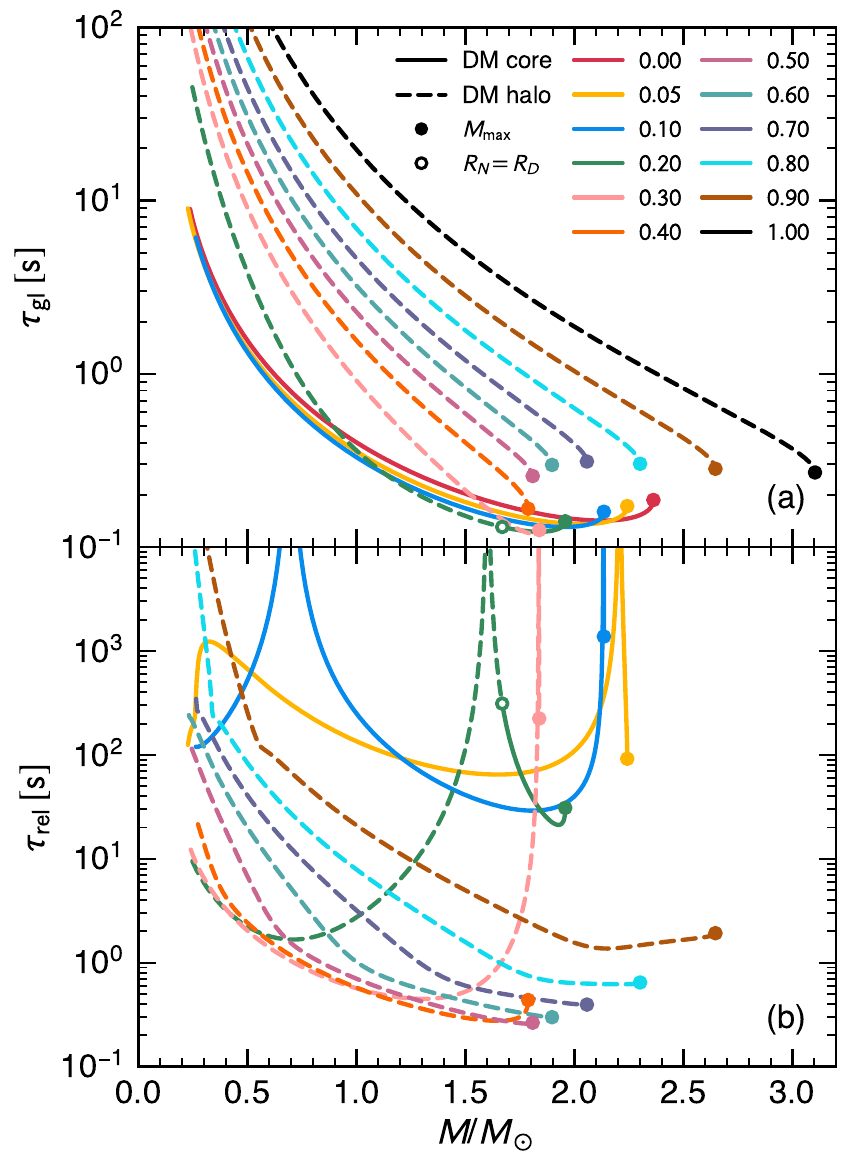}
\caption{
Full-GR damping time $\tau=1/\om_I$ for the two principal $f$-like sequences.
The vertical axis is logarithmic.
Panels (a) and (b) show $\fgl$ and $\frel$, respectively.
Colors label the fixed DM mass fraction $\fdm$,
with the legend giving the corresponding values,
and filled circles mark the maximum-mass configurations.
The tallest narrow peaks in the $\frel$ panel exceed the displayed vertical range,
and representative numerical values are quoted in the text.
}
\label{f:tf_tau}
\end{figure}

The full-GR damping times are shown in Fig.~\ref{f:tf_tau}.
In panel (a), the ordinary-like part of $\fgl$
has a damping time typically of the same order as ordinary NS $f$ modes,
with values around fractions of a second to several seconds over much of the mass range.
The increase of $\tau$ toward low mass is expected,
because the compactness and radiative efficiency
decrease as the star becomes more diffuse.
The pure ordinary sequence has $\tau\simeq0.14$--$9~{\rm s}$
in the plotted range.
For DM-rich global branches,
the low-mass configurations can have much longer damping times,
reaching $10^2$--$10^3~{\rm s}$,
mainly because the configurations are extended and radiate inefficiently.
Near the maximum-mass points,
however, the global-like,
non-cancelling branches typically return to $\tau\sim0.1$--$0.4~{\rm s}$.
The same compactness trend is seen in ordinary full-GR $f$-mode calculations.
In both cases,
the damping time generally varies smoothly with compactness
along a one-fluid-like sequence \cite{Sotani21,Kunjipurayil22,Pradhan22}.
In that setting a shorter damping time
usually indicates more efficient emission of GWs
from a compact configuration.

Panel (b) shows that the $\frel$ branch exhibits a much wider range of damping times,
including narrow regions where $\tau$ grows by several orders of magnitude.
These long-lived points correspond to weak coupling to the outgoing radiative channel.
In many cases they coincide with the low-$R_Q$ regions in Fig.~\ref{f:tf_diag},
supporting the interpretation that the ordinary-matter and DM quadrupoles nearly cancel.
The correspondence is not required to be exact,
because $R_Q$ is a PN-motivated matter-source diagnostic,
whereas $\tau$ is determined by the complete exterior outgoing Zerilli condition.
Metric perturbations, surface matching, pressure and momentum perturbations,
and propagation through the exterior potential
all enter the radiative projection.
Therefore a large damping time indicates weak full-GR mode-to-wave coupling.
Matter-quadrupole cancellation is one way to produce it,
but the outgoing Zerilli amplitude depends on metric,
surface-matching,
and exterior-propagation effects
that are not contained in $R_Q$.
The $\fdm=0.10$ $\frel$ sequence contains two prominent damping-time peaks,
near $M\simeq0.71\ms$ and $M\simeq2.13\ms$, with
$\tau\simeq2.8\times10^7~{\rm s}$ and
$\tau\simeq6.5\times10^7~{\rm s}$,
respectively.
For comparison, the largest values
found on the $\fdm=0.2$ and $0.3$ sequences
are approximately
$1.2\times10^6~{\rm s}$ near $M\simeq1.60\ms$ and
$6.2\times10^6~{\rm s}$ near $M\simeq1.84\ms$.
For the most weakly damped models,
the imaginary part of the eigenfrequency
can approach the absolute accuracy of the complex root search.
The peak heights and their precise locations
should therefore be regarded as approximate;
the more robust result is the presence of narrow weakly radiating regions
and the associated suppression of the radiative diagnostics.
The low-mass $\fdm=0.10$ peak is close to a low-$R_Q$ region,
with $C_{ND}\simeq-1$,
whereas the high-mass peak in Table~\ref{t:weak_diag}
is less directly tied to a minimum of $R_Q$.
Its matter quadrupole is reduced but not nearly zero.
Such cases emphasize that the full damping time is
controlled by the outgoing Zerilli amplitude,
not by the matter quadrupole diagnostic alone.
The mode can remain a fluid-dominated oscillation in the interior,
while its overlap with the radiative exterior solution is strongly suppressed.

The contrast is summarized in Table~\ref{t:weak_diag}.
The low-mass point has $R_Q\ll1$,
showing that the matter quadrupole itself is nearly cancelled.
Its metric response is also small throughout the star.
The high-mass point is different:
the matter quadrupole is not close to zero,
and the internal metric perturbation reaches a much larger amplitude.
It remains sizable at the inner DM surface,
but it is suppressed by more than four orders of magnitude
at the outer ordinary-matter surface.
The normalized Zerilli-wave luminosity measure $\mathcal{R}_Z$
is also much smaller.
Thus the weak radiation in this case
is not caused by a small internal metric perturbation everywhere.
Instead,
the suppression is associated with a small surface Zerilli combination,
and hence with a small projection onto the outgoing wave amplitude $A_Z$.
Absolute comparisons between the two rows are not physical
because the two eigenfunctions are normalized independently.

In the relevant parts of parameter space,
the two-fluid star therefore supports a bright combination,
in which the component quadrupoles add with little cancellation,
and a weak combination,
in which the relative fluid degree of freedom
suppresses the radiative projection.
The latter can store oscillation energy in the fluid motion
and near-zone metric deformation
while emitting very little energy to infinity.
The corresponding one-fluid problem
lacks this additional composition-space direction,
although radial source cancellation can still occur
in modes with nodes.

\subsection{Energy required for a detectable signal}

\begin{figure*}[t]
\centering
\includegraphics[width=0.83\textwidth]{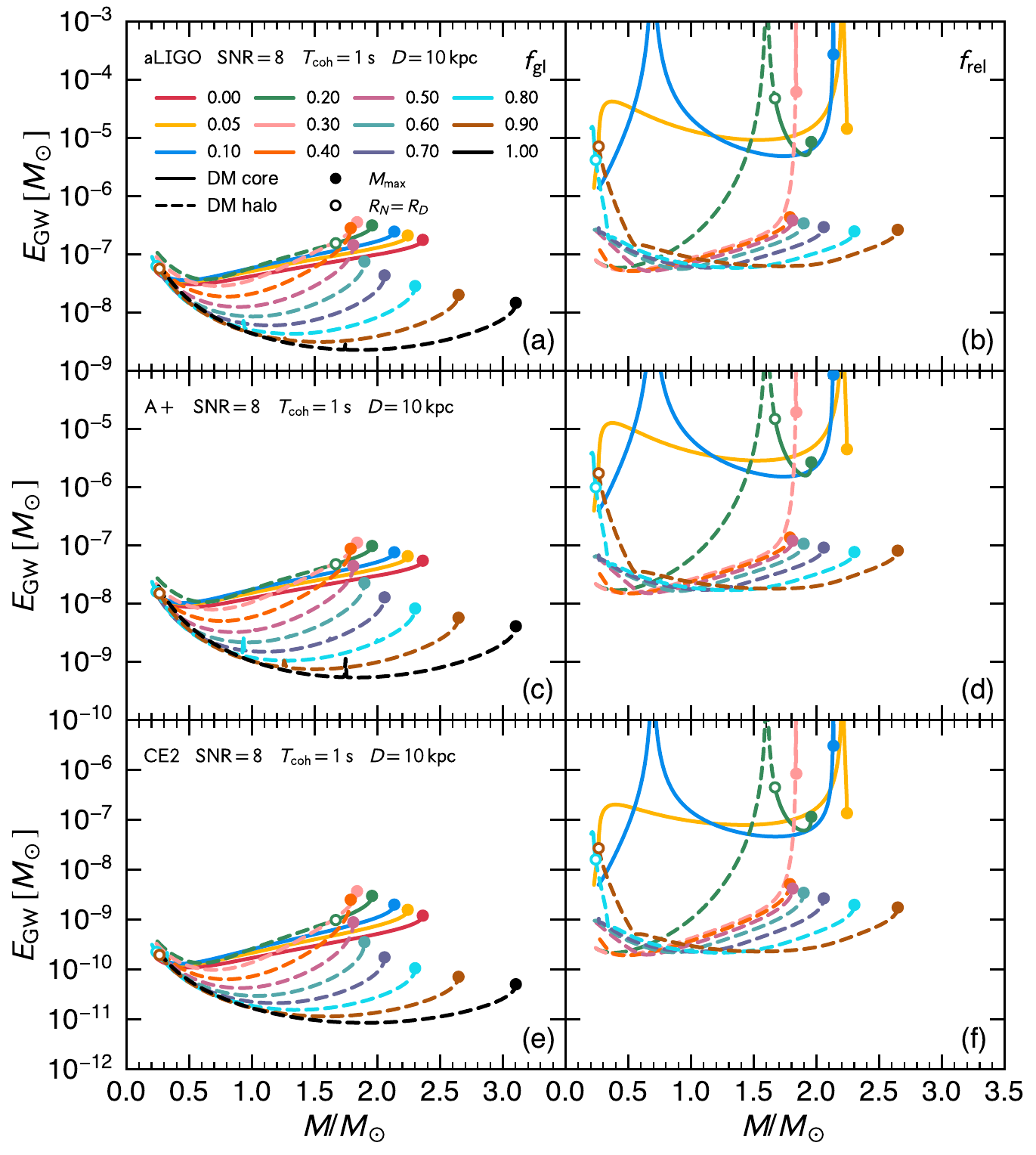}
\caption{
GW energy required to reach a matched-filter threshold ${\rm SNR}=8$
for a source at $D=10~{\rm kpc}$,
assuming a coherent integration time $T_{\rm coh}=1~{\rm s}$.
The energy is quoted in units of $M_\odot$.
Panels (a) and (b), (c) and (d), and (e) and (f) correspond to
aLIGO, A+, and CE2 sensitivities, respectively
\cite{Abbott18b,Aplus,CE1ET,Hild11}. The left column, panels (a),
(c), and (e), shows $\fgl$, and the right column, panels (b), (d),
and (f), shows $\frel$.
Colors label the fixed DM mass fraction $\fdm$,
with the legend giving the corresponding values.
Filled circles mark the maximum-mass configurations.
The curves include the finite-time factor in Eq.~(\ref{e:snr_narrow}).
}
\label{f:tf_ereq}
\end{figure*}

Figure~\ref{f:tf_ereq} translates the mode frequencies and damping times
into the GW energy required to reach the fixed matched-filter threshold ${\rm SNR}=8$.
This is the standard damped-sinusoid estimate
used in GW asteroseismology studies of isolated oscillating stars \cite{Benhar04,Pradhan22},
with the full-GR damping time and the finite coherent integration factor
included through Eq.~(\ref{e:snr_narrow}).
The source distance is set to $D=10~{\rm kpc}$,
and the coherent integration time is fixed to $T_{\rm coh}=1~{\rm s}$.
The vertical axis is the total GW energy expressed
in the solar-mass energy unit $M_\odot$.
The calculation uses the narrow-band expression Eq.~(\ref{e:snr_narrow}),
with the detector noise evaluated at the mode frequency.
The plotted quantity therefore includes
the effects of the detector sensitivity curve,
the mode frequency,
the damping time,
and the finite coherent observation time.

For the finite coherent time adopted here,
segments with $\tau\gg T_{\rm coh}$ generally
require more total radiated energy
than segments with comparable frequency
and $\tau\lesssim T_{\rm coh}$ to reach the same SNR.
This trend follows from the finite-time factor in Eq.~(\ref{e:snr_energy});
the mode frequency and detector noise remain equally important.
In the fully coherent limit $T_{\rm coh}\gg\tau$,
the required radiated energy becomes independent of the damping time.
For the next-generation CE2 curve in panels (e) and (f),
the required energy can fall many orders of magnitude
below that for aLIGO,
as expected from the improved high-frequency sensitivity.
For the bright part of $\fgl$ in panels (a), (c), and (e),
the minima are typically around
$10^{-8}$--$10^{-7}\,M_\odot$ for aLIGO,
$10^{-9}$--$10^{-7}\,M_\odot$ for A+,
and $10^{-11}$--$10^{-9}\,M_\odot$ for CE2.
The exact value moves along the sequence
because the detector noise is sampled at the mode frequency.
This produces the broad minima in panels (a), (c), and (e).
They occur where the mode is sufficiently radiative,
where the damping time is not much longer than the coherent time,
and where the frequency lies in a comparatively favorable part
of the detector noise curve.
The weakly radiating segments, however,
can become difficult to detect
even when their damping times are very large.
At fixed total radiated energy,
the initial strain scales as $h_0\propto\tau^{-1/2}$.
If $\tau$ exceeds the time over which the signal can be coherently tracked,
the SNR gains only the factor $\sqrt{T_{\rm coh}}$ rather than $\sqrt{\tau/2}$.
Thus long-lived weakly radiating modes do not automatically yield a large SNR.
Their detectability is controlled by the radiated power per unit time,
the frequency location relative to the detector noise,
the distance, the angular projection,
and the available coherent integration time.
In panels (b), (d), and (f), this produces sharp
increases of the required energy near the same masses
where $\tau$ is strongly enhanced.
Some of these peaks run to the upper edge of the plotted range.
The sharp peaks are radiative inefficiency markers.
They mark configurations where a large internal oscillation energy
would be needed to produce the same detector-frame SNR.

The required-energy peaks occur at the same frequencies
and masses where the radiative channel is suppressed.
A weakly radiating segment may therefore be more useful
as a diagnostic of two-fluid internal dynamics
than as a favorable GW target.
If it were observed together with a bright companion mode,
the combination of frequency splitting
and very different damping times
would provide a distinctive signature
of an additional gravitationally coupled fluid component.
The curves instead quantify the GW energy
that would be required under the stated detector,
distance, orientation, and coherence assumptions;
they do not predict the energy deposited into either branch
by a merger, glitch, or collapse.

\section{Discussion}
\label{s:disc}

The two sequences found here
cannot be identified simply by the component
that carries most of the kinetic energy.
The $\fgl$ sequence connects continuously
to the ordinary single-fluid $f$ mode as $\fdm\to0$,
whereas $\frel$ is associated with the additional relative degree of freedom.
Along either sequence,
however, the dominant fluid component
and the relative phase can change with mass and DM fraction.
The labels $\fgl$ and $\frel$
should therefore be understood as continuation labels
rather than fixed physical identifications.

The comparison with previous Cowling studies
clarifies the role of metric-mediated coupling.
In the Cowling approximation,
the two fluid perturbations are independent
at linear order on the same two-fluid equilibrium background.
The resulting frequencies characterize component-associated fluid responses,
rather than coupled radiative normal modes \cite{Sotani25a,Sotani25b}.
The approximation therefore omits both the perturbed spacetime
and the outgoing-wave boundary condition.
Ref.~\cite{Kumar26} formulated
the corresponding coupled full-GR perturbation problem.
Here we use that formulation to compute the complex eigenfrequency,
the GW damping time,
the asymptotic Zerilli amplitude,
and the detector-oriented energy estimates.
Consequently,
the Cowling frequencies can be used as qualitative guides,
but they do not by themselves determine the radiative character of either branch.

The weakly radiating configurations fall
into two physically different categories.
In the first,
the ordinary-matter and DM quadrupoles
are nearly opposite and comparable,
so that $R_Q\ll1$.
This interpretation is supported
when $C_Q\simeq-1$ and $|Q_{\dm}/Q_{\nm}|\simeq1$.
In the second,
the total matter quadrupole is not close to zero,
but the surface perturbation has only a small projection
onto the outgoing Zerilli solution.
The high-mass $\fdm=0.10$ example
in Table~\ref{t:weak_diag} illustrates this second case.
Its weak radiation is therefore controlled
by the surface-matched radiative amplitude,
rather than by $R_Q$ alone.

The largest values of $\tau$ should be interpreted with care,
because the corresponding $\om_I$ can approach
the absolute tolerance of the complex root search.
The existence of weakly radiating regions
is more robust than the precise height or location of the narrowest damping-time peaks.
Convergence with respect to the root tolerance,
exterior matching, and radial integration
should therefore be checked before assigning high precision to those peaks.
The present sequences also use one ordinary-matter EOS,
one DM EOS, fixed DM mass fractions, and nonrotating backgrounds.
Extending the calculation to additional EOSs, rotation,
and physically motivated mode-excitation mechanisms is left for future work.

\section{Conclusions}
\label{s:end}

We have solved the quadrupolar polar perturbation problem
of gravitationally coupled two-fluid stars in full GR,
including the outgoing exterior boundary condition.
This yields complex eigenfrequencies, damping times,
and asymptotic Zerilli amplitudes along fixed-$\fdm$ sequences.
The corresponding equilibrium sequences
change from compact DM-core configurations
to extended DM-halo stars as the DM fraction increases.

The spectrum contains two principal $f$-like continuations.
The $\fgl$ branch connects to the ordinary single-fluid $f$ mode,
whereas $\frel$ is associated
with the additional relative degree of freedom.
Neither branch has a fixed NM- or DM-led interpretation over the full sequence;
the local character is determined from the component energies,
displacement overlap, quadrupole diagnostics, and exterior-wave quantities.

Weakly radiating regions occur on the $\frel$ continuation.
Some are caused by nearly opposite and comparable matter quadrupoles,
identified by small $R_Q$ together with $C_Q\simeq-1$.
Others have a non-negligible matter quadrupole
but a strongly suppressed surface projection
onto the outgoing Zerilli solution.
The outgoing amplitude and the damping time
are therefore the direct radiative diagnostics,
while $R_Q$ identifies only one source-level mechanism.

Finally,
we translated the modes into the GW energy required
to reach a fixed matched-filter SNR.
Long damping times do not by themselves make a mode easier to detect:
for fixed radiated energy the initial strain scales as $\tau^{-1/2}$,
and a finite coherent integration time limits the SNR gain.
The observational relevance of a branch
therefore depends on the radiated power,
the detector noise at the mode frequency,
the distance and angular projection,
and the available coherence time.
Future work should extend the calculation
to broader nuclear and dark-sector EOS sets,
include rotation,
and connect the mode amplitudes
to dynamical excitation mechanisms
and detector-network analyses.

\begin{acknowledgments}

Zi-Yue Zheng and Xiao-Ping Zheng are supported by
the National Natural Science Foundation of China (Grant No.~12473039).
Jin-Biao Wei and Huan Chen acknowledge financial support
from the National Natural Science Foundation of China (Grant No.~12205260).

\end{acknowledgments}

\newcommand{\epja}{Euro. Phys. J. A}
\newcommand{\aap}{Astron. Astrophys.}
\newcommand{\apjl}{Astrophys. J. Lett.}
\def\jcap{Journal of Cosmology and Astroparticle Physics}
\def\jcap{JCAP}
\newcommand{\mnras}{Mon. Not. R. Astron. Soc.}
\newcommand{\nphysa}{Nucl. Phys. A}
\newcommand{\physrep}{Phys. Rep.}
\newcommand{\plb}{Phys. Lett. B}
\newcommand{\ppnp}{Prog. Part. Nucl. Phys.}
\bibliography{dansnro}

\end{document}